\documentclass[reprint,superscriptaddress,nofootinbib,amsmath,amssymb,aps]{revtex4-2}
\usepackage{graphicx}
\usepackage{float}
\usepackage{dcolumn}
\usepackage{bm}
\usepackage{hyperref}
\usepackage{tikz}
\usetikzlibrary{quantikz}

\begin{document}

\title{Quantum computing of chirality imbalance in SU(2) gauge theory}

\author{Guofeng Zhang }
 \affiliation{Key Laboratory of Atomic and Subatomic Structure and Quantum Control (MOE), Guangdong Basic Research Center of Excellence for Structure and Fundamental Interactions of Matter, Institute of Quantum Matter, South China Normal University, Guangzhou 510006, China}
\affiliation{Guangdong-Hong Kong Joint Laboratory of Quantum Matter, Guangdong Provincial Key Laboratory of Nuclear Science, Southern Nuclear Science Computing Center, South China Normal University, Guangzhou 510006, China }
\author{Xingyu Guo }
 \email{guoxy@m.scnu.edu.cn}
 \affiliation{Key Laboratory of Atomic and Subatomic Structure and Quantum Control (MOE), Guangdong Basic Research Center of Excellence for Structure and Fundamental Interactions of Matter, Institute of Quantum Matter, South China Normal University, Guangzhou 510006, China}
\affiliation{Guangdong-Hong Kong Joint Laboratory of Quantum Matter, Guangdong Provincial Key Laboratory of Nuclear Science, Southern Nuclear Science Computing Center, South China Normal University, Guangzhou 510006, China }

\author{Enke Wang }
 \email{wangek@scnu.edu.cn}
 \affiliation{Key Laboratory of Atomic and Subatomic Structure and Quantum Control (MOE), Guangdong Basic Research Center of Excellence for Structure and Fundamental Interactions of Matter, Institute of Quantum Matter, South China Normal University, Guangzhou 510006, China}
\affiliation{Guangdong-Hong Kong Joint Laboratory of Quantum Matter, Guangdong Provincial Key Laboratory of Nuclear Science, Southern Nuclear Science Computing Center, South China Normal University, Guangzhou 510006, China }

\author{Hongxi Xing }%
 \email{hxing@m.scnu.edu.cn}
 \affiliation{Key Laboratory of Atomic and Subatomic Structure and Quantum Control (MOE), Guangdong Basic Research Center of Excellence for Structure and Fundamental Interactions of Matter, Institute of Quantum Matter, South China Normal University, Guangzhou 510006, China}
\affiliation{Guangdong-Hong Kong Joint Laboratory of Quantum Matter, Guangdong Provincial Key Laboratory of Nuclear Science, Southern Nuclear Science Computing Center, South China Normal University, Guangzhou 510006, China }
\affiliation{Southern Center for Nuclear-Science Theory (SCNT),
Institute of Modern Physics, Chinese Academy of Sciences, Huizhou 516000, China}

\collaboration{QuNu Collaboration}

\date{\today}

\begin{abstract}
We implement a variational quantum algorithm to investigate the chiral condensate in a 1+1 dimensional SU(2) non-Abelian gauge theory. The algorithm is evaluated using a proposed Monte Carlo sampling method, which allows the extension to large qubit systems. The obtained results through quantum simulations on classical and actual quantum hardware are in good agreement with exact diagonalization of the lattice Hamiltonian, revealing the phenomena of chiral symmetry breaking and restoration as functions of both temperature and chemical potential. Our findings underscore the potential of near-term quantum computing for exploring QCD systems at finite temperature and density in non-Abelian gauge theories.
\end{abstract}

\maketitle

\section{\label{sec:level1}Introduction\protect}
Spontaneous chiral symmetry breaking is a fundamental phenomenon in Quantum Chromodynamics (QCD), which governs the strong interactions between quarks and gluons \cite{Weinberg:1995mt}. The chiral condensate, formed by the quark-antiquark pairs due to spontaneous chiral symmetry breaking, imparts a non-zero expectation value to the bilinear fermion operator $\bar{\psi}\psi$, which is crucial for comprehending the origin of quark mass \cite{Gell-Mann:1968hlm,Nambu:1961tp,Nambu:1961fr,Goldstone:1961eq,Shifman:1978bx,Weinberg:1975ui}. In QCD systems with high temperature, quarks and gluons deconfine, leading to the formation of quark-gluon plasma (QGP) and chiral symmetry restoration \cite{Pisarski:1983ms}. On the other hand, under high-density conditions, quarks are highly overlapping, and the ordinary quark-antiquark condensates are replaced by color-flavor locking of quark Cooper pairs \cite{Alford:1998mk,Rapp:1997zu}. The restoration of chiral symmetry at high temperature and density is crucial to understanding the evolution of the early universe, as well as the properties of QGP and neutron stars \cite{Alford:2007xm,Rajagopal:2000wf,Stephanov:1998dy,Fukushima:2010bq}.

Due to the inherently non-perturbative nature of QCD, the first principle theoretical and numerical studies of chirality imbalance present significant challenges \cite{Gross:1973id,Politzer:1973fx}. For finite temperature systems, lattice gauge theory (LGT) has emerged as a prominent theory in addressing these challenges\cite{Wilson:1974sk,Kogut:1979wt}.
In LGT, as constrained by gauge invariance to facilitate quark-gluon interactions, the matter degrees of freedom are encoded into the lattice sites and the gauge degrees of freedom are represented by the links between different lattice sites. Monte Carlo methods, employed in LGT, provide a direct approach for solving non-perturbative QCD problems from first principles and have achieved significant success \cite{Fukushima:2010bq,Aoki:2016frl,BMW:2008jgk}. However, such a method encounters the notorious sign problem \cite{Troyer:2004ge} when addressing issues related to finite chemical potential for high density systems. Though Taylor expansion techniques have been performed to alleviate the sign problem for small chemical potential systems \cite{deForcrand:2002hgr}, the issue in high chemical potential systems remains unresolved. 

In recent years, with the rapid development of quantum hardware, quantum computing has become increasingly prominent in dealing with the sign problems in high-energy physics \cite{Bauer:2022hpo,DiMeglio:2023nsa,Zhang:2020uqo,Fang:2024ple}, such as nucleon structure \cite{Li:2021kcs,Li:2022lyt,Li:2023kex,Lamm:2019uyc} and parton fragmentation \cite{Li:2024nod,Grieninger:2024axp}. In addition, another major advantages of quantum simulation is that it enables the study of thermal equilibrium systems while circumventing the sign problem associated with traditional Monte Carlo methods \cite{Czajka:2021yll,deJong:2021wsd,Yang:2020yer,Zhou:2021kdl,Kokail:2018eiw,Xie:2022jgj,Davoudi:2022uzo}. In particular, quantum simulations for real-time chiral dynamics at finite temperature have been investigated in 1+1 dimensional Nambu-Jona-Lasinio model \cite{Czajka:2021yll}, SYK model\cite{Araz:2024xkw} and Schwinger model \cite{Ikeda:2024rzv} using quantum imaginary time evolution algorithm\cite{Motta:2019yya}. 

In this paper, we investigate the chiral symmetry breaking and restoration in finite-temperature and density QCD systems. As a first step considering non-Abelian gauge theory, we take SU(2) gauge theory to demonstrate the simulation. SU(2) exhibits several physical characteristics similar to QCD, such as spontaneous chiral symmetry breaking, quark confinement, asymptotic freedom, and so on \cite{RevModPhys.55.775,tHooft:1998ifg}. The quantum simulations of SU(2) have been successfully realized for various problems, such as hadron spectroscopy \cite{Atas:2021ext}, ground state and energy gaps \cite{Ale:2024uxf}, as well as shear viscosity \cite{Turro:2024pxu}.

We simulate explicitly the chiral condensate in finite temperature and density through the evaluation of order parameter for spontaneous chiral symmetry breaking, using the 1+1 dimensional SU(2) non-Abelian gauge theory. To obtain the thermal average of the chiral condensate, it is essential to efficiently prepare the Gibbs state on a quantum computer. Given that the Gibbs state corresponds to the minimum of the free energy, the strategy involves preparing the Gibbs state is minimizing the free energy by variational methods. The Quantum Alternating Operator Ansatz (QAOA) \cite{Farhi:2014ych}, known for its robust representation power, is utilized to parameterize the free energy. The Monte Carlo method is commonly used in lattice field theory to simulate thermal systems effectively. In particular, we implement the Monte Carlo random sampling for the optimization of eigenstates at a specific temperature, which also suffices for any other temperatures, significantly reducing the computational cost. Additionally, the Monte Carlo method is used to calculate the desired density matrix, offering a substantial advantage for large qubit systems. We observe spontaneous chiral symmetry breaking at low temperatures and automatic restoration at high temperatures within the non-Abelian theoretical framework by variational quantum algorithm, which is consistent with theoretical expectations. We also investigate the behavior of chiral condensate at different chemical potentials. To validate the effectiveness of our variational method, we compare our simulation results with those obtained from exact diagonalization. Our work demonstrates the potential of using near-term quantum computers to explore QCD systems with finite temperature and finite density.

The remainder of this paper is organized as follows. We first introduce the 1+1 dimensional SU(2) gauge theory, and briefly illustrate how to eliminate gauge fields and obtain the qubit formulation in Sec. \ref{sec:level2}. Then, we propose a variational quantum algorithm to calculate the thermal average of chiral condensate. We present the simulation results in Sec. \ref{sec:level3} and summarize in Sec. \ref{sec:level4}.

\section{Methods\label{sec:level2}}
In this section, we provide a brief introduction to the lattice SU(2) gauge theory using staggered fermions. We illustrate how to eliminate the gauge fields in the kinetic terms and apply Gauss's law to represent the chromoelectric field with fermionic operators. In the end, the Jordan-Wigner transformation is introduced to map the fermionic degrees of freedom to the Pauli spin system. 

\subsection{SU(2) non-Abelian gauge theory on the lattice}

We start with the 1+1 dimensional Yang-Mills gauge theory, whose Lagrangian density in continuous spacetime\cite{Taylor1979GaugeTO} is given by\footnote{Our convention:$\eta_{\mu\nu}=\text{diag}(1,-1),\gamma^{0}=\sigma_{z},\gamma^{1}=i\sigma_{y},\gamma^{5}=\sigma_{x},\\F_{\mu\nu}^{a}=\partial_{\mu}A_{\nu}^{a}-\partial_{\nu}A_{\mu}^{a}+gf^{abc}A_{\mu}^{b}A_{\nu}^{c},f^{abc}=-2i\text{Tr}\left(\left[t^{a},t^{b}\right],t^{c}\right)$}
\begin{eqnarray}
\mathcal{{L}}=-\frac{1}{4}F_{\mu\nu}^{a}F^{\mu\nu,a}+i\bar{\psi}\gamma^{\mu}(\partial_{\mu}+igA_{\mu}^{a}t^{a})\psi-m\bar{\psi}\psi,
    \label{eq:a}
\end{eqnarray}
where $\psi = \psi^{c = r,g}_{\alpha = 1,2}$ is a four-component fermionic operator. The $A_{\mu}^{a}t^{a}$ represent non-Abelian gauge potential, where $t^{a}$ are the three generators of the Lie algebra associated with the SU(2) group, and $a$ indicates the index of the Pauli operators. The parameter $m$ represents the bare mass of the fermion, while $g$ describes the interaction strength between the fermionic matter field and the gauge field, as well as the self-coupling strength of the gauge field.

Adopting the temporal gauge $A_{0}^{a}=0$, the Hamiltonian density of the continuum theory is found to be
\begin{eqnarray}
H=- i\bar{\psi} \gamma^1 (\partial_1 + igA_1^{a}t^{a}) \psi + m\bar{\psi}\psi+\frac{1}{2} \left( L^{a} \right)^2,
\label{eq:b}
\end{eqnarray}
where the chromoelectric field $L^{a}=-\partial^{0}A^{1,a}$ serves as the canonical conjugate momentum of $A_{1}^{a}$, satisfying the commutation relation 
$\left[A_{1}^{a}(x),L^{b}(y)\right]=i\delta^{ab}\delta(x-y)$.

For the purposes of facilitating quantum simulation, we discretize the continuous version of the Hamiltonian using staggered fermions. The Kogut-Susskind formulation of the Yang-Mills Hamiltonian\cite{Kogut:1979wt,Kogut:1974ag}\cite{Zohar:2014qma,Grabowska:2024emw}, is given by
\begin{eqnarray}
H=&&\frac{1}{2\Delta}\sum_{n=0}^{N-2}\left(\phi_{n}^{\dagger}U_{n}\phi_{n+1}+H.C.\right)\nonumber\\
&&+m\sum_{n=0}^{N-1}(-1)^{n+1}\phi_{n}^{\dagger}\phi_{n}+\frac{\Delta g^{2}}{2}\sum_{n=0}^{N-2}\mathbf{L}_{n}^{2},
\label{eq:c}
\end{eqnarray}
where $\Delta$ is the lattice spacing, $N$ denotes the number of lattice sites. The field $\phi_{n}=\left(\phi_{n}^{r},\phi_{n}^{g}\right)^{T}$ represents the staggered fermion field defined at lattice site $n$, with $\phi_{n}^{r}$ and $\phi_{n}^{g}$ corresponding to its red and green color components. The four components of a Dirac fermion are replaced by a pair of neighboring staggered fermions defined by
\begin{eqnarray}
\frac{\phi_n}{\sqrt{\Delta}} \longleftrightarrow
\begin{cases}
\psi_1(x) & n:\text{even}, \\
\psi_2(x) & n:\text{odd}.
\end{cases}
\label{eq:d}
\end{eqnarray}
In Eq. (\ref{eq:c}), the gauge link $U_{n}=$ exp$\left(i\theta_{n}^{a}t^{a}\right)$, extending from lattice site $n$ to lattice site $n+1$, ensures that the Hamiltonian is invariant under local gauge transformations. The non-Abelian phase $\theta_{n}^{a}$ is related to the gauge field as $\theta_{n}^{a}=-agA_{n}^{1,a}$.

The last term of the Hamiltonian represents the energy of the chromoelectric field, in which the chromoelectric field $\mathbf{L}_{n}$ can be expressed through its left chromoelectric field or right chromoelectric field components, $\mathbf{L}_{n}^{2}=L_{n}^{a}L_{n}^{a}=R_{n}^{a}R_{n}^{a}$. The left and right chromoelectric fields are related through the adjoint representation of the SU(2) group,  $R_{n}^{a}=\left(U_{n}^{{\rm adj}}\right)^{ab}L_{n}^{b}$\footnote{The matrix element of the SU(2) group in the adjoint representation is given by: $(U_{n}^{{\rm adj}})^{ab}=2{\rm Tr}[U_{n}t^{a}U_{n}^{\dagger
}t^{b}]$}, and satisfy the algebra
\begin{eqnarray}
&&\left[R_{n}^{a},R_{m}^{b}\right]=i\epsilon^{abc}R_{n}^{c}\delta_{mn},\nonumber\\
&&\left[L_{n}^{a},L_{m}^{b}\right]=-i\epsilon^{abc}L_{n}^{c}\delta_{mn},\nonumber\\
&&\left[L_{n}^{a},R_{m}^{b}\right]=0.
\label{eq:e}
\end{eqnarray}
As the conjugate momenta of the gauge field, the chromoelectric fields and the gauge link satisfy the canonical commutation relation
\begin{eqnarray}
&&\left[L_{n}^{a},\left(U_{n}\right)_{\alpha\beta}\right]=\left(t^{a}U_{n}\right)_{\alpha\beta},\nonumber\\
&&\left[R_{n}^{a},\left(U_{n}\right)_{\alpha\beta}\right]=\left(U_{n}t^{a}\right)_{\alpha\beta}.
\label{eq:f}
\end{eqnarray}

The physical state $\Psi$ must satisfy Gauss's law, which is expressed as $\mathbf{G}_{n} |\Psi\rangle=0$ where $G$ is the Gauss operator. This implies that the eigenstates of the Gauss operator span the Hilbert space. The three components of the Gauss operator can be represented as $G_{n}^{a}=L_{n}^{a}-R_{n-1}^{a}-Q_{n}^{a}$, where the non-Abelian charge $Q_{n}^{a}$ consists of static and dynamic charges. The dynamic charges are defined as
\begin{eqnarray}
Q_{n}^{a}=\phi_{n}^{\dagger}t^{a}\phi_{n}, \quad a=x,y,z.
\label{eq:g}
\end{eqnarray}

\subsection{Purely fermionic formulation and qubit encoding}
For the 1+1 dimensional SU(2) non-Abelian gauge theory, the gauge degrees of freedom are not completely independent, which allows the Kogut-Susskind Hamiltonian to be expressed in purely fermionic form. We briefly describe how local gauge transformations, combined with Gauss's law and possible boundary conditions \cite{Hamer:1976bj,Bringoltz:2008iu,Lenz:1994cv,Sala:2018dui} %(by specifying the incoming chromoelectric field flux to determine chromoelectric field excitations throughout the lattice), 
eliminate all the gauge operators in the Hamiltonian. 

Before proceeding, for clarity in representation, we rewrite the Hamiltonian as
\begin{eqnarray}
H=\tilde{m}H_{\text{m}}+\frac{\Delta^2 g^{2}}{2}H_{\text{el}}+\frac{1}{2}H_{\text{kin}},
\label{eq:h}
\end{eqnarray}
where $\tilde{m}=\Delta m$.

To eliminate the gauge links, we consider the following local gauge transformation 
\begin{eqnarray}
\Theta = \prod_{k} \exp\left(i \mathbf{\theta}_k \cdot \sum_{j>k} \mathbf{Q}_j\right).
\label{eq:i}
\end{eqnarray}
Under this transformation, the mass term remains the
same, and the hopping term becomes
\begin{eqnarray}
H_{\text{kin}}^{\prime}\rightarrow\Theta H_{\text{kin}}\Theta^{\dagger}=\sum_{n=0}^{N-2}(\phi_{n}^{\dagger}\phi_{n+1}+\text{H.C.}).
\label{eq:j}
\end{eqnarray}
Consequently, the Hamiltonian specified in Eq. \eqref{eq:c} can be reformulated as
\begin{eqnarray}
H^{\prime} \rightarrow\Theta H \Theta^\dagger &&= \frac{1}{2} \sum_{n=0}^{N-2} \left( \phi_n^\dagger \phi_{n+1} + {\rm H.C.} \right) \nonumber\\
&&+ \tilde{m} \sum_{n=0}^{N-1} (-1)^{n+1} \phi_n^\dagger \phi_n + \frac{\Delta^2 g^2}{2}  H_{\rm el}^{\prime}.
\label{eq:k}
\end{eqnarray}

Now we implement Gauss's law by considering open boundary condition, where the incoming chromoelectric field flux is set as $\mathbf{R}_{-1}=0$, then the energy term of the chromoelectric field can be expressed in terms of the fermionic field 
\begin{eqnarray}
H_{\text{el}}^{\prime}\rightarrow\Theta H_{\text{el}}\Theta^{\dagger}&&=\sum_{n=0}^{N-2}\Theta\mathbf{L}_{n}^{2}\Theta^{\dagger}=\sum_{n=0}^{N-2}\left(\sum_{k\leq n}\mathbf{Q}_{k}\right)^{2}.\quad
\label{eq:l}
\end{eqnarray}
The methodology presented above allows us to eliminate the gauge field in the Hamiltonian, but with the cost of introducing long-range interactions in the chromoelectric field energy term.

In order to work on digital quantum computers, one needs to encode the Hamiltonian in terms of a spin system. We employ the standard Jordan-Wigner transformation \cite{Jordan1928berDP} to map the fermionic matter degrees of freedom to Pauli spin operators
\begin{equation}
\varphi_{n}=\sigma^{-}_{n}\prod_{l=0}^{n-1}\left(-i\sigma_{l}^{z}\right),
\label{eq:m}
\end{equation}
where $\sigma^{\pm}=\frac{1}{2}(\sigma^{x}\pm i\sigma^{y})$, we replace the two components of the staggered fermion with
\begin{eqnarray}
\phi_{n}=
\begin{pmatrix}
\phi^{r}_{n} \\ 
\phi^{g}_{n} 
\end{pmatrix}
\longleftrightarrow
\begin{pmatrix}
\varphi_{2n} \\ 
\varphi_{2n+1}
\end{pmatrix}.
\label{eq:n}
\end{eqnarray}
Therefore, the lattice Hamiltonian can be mapped to the spin system as follows 
\begin{eqnarray}
H=&&\tilde{m}\sum_{n=0}^{N-1}\left[\left(-1\right)^{n+1}\frac{\sigma_{2n}^{z}+\sigma_{2n+1}^{z}}{2}+1\right]\nonumber\\
&&-\frac{1}{2}\sum_{n=0}^{N-2}\left(\sigma_{2n}^{+}\sigma_{2n+1}^{z}\sigma_{2n+2}^{-}+\sigma_{2n+1}^{+}\sigma_{2n+2}^{z}\sigma_{2n+3}^{-}+\text{H.C.}\right)\nonumber\\
&&+\frac{\Delta^2 g^{2}}{2}\sum_{n=0}^{N-2}\left(\sum_{k\leq n}\mathbf{Q}_{k}\right)^{2},
\label{eq:o}
\end{eqnarray}
where the three components of the non-Abelian charge are encoded as
\begin{eqnarray}
&&Q_{n}^{x}=\frac{1}{2}\left(\sigma_{2n+1}^{+}\sigma_{2n}^{-}+\text{H.C.}\right)\nonumber\\
&&Q_{n}^{y}=\frac{i}{2}\left(\sigma_{2n+1}^{+}\sigma_{2n}^{-}-\text{H.C.}\right)\nonumber\\
&&Q_{n}^{z}=\frac{1}{4}\left(\sigma_{2n}^{z}-\sigma_{2n+1}^{z}\right).
\label{eq:p}
\end{eqnarray}

Moreover, for finite density system in SU(2) non-Abelian gauge theory, the Hamiltonian can be expressed by adding an extra term $-\mu\sum_{n=0}^{N-1}\phi_{n}^{\dagger}\phi_{n}$ \cite{Lowdin:1955zzb,Lowdin:1955zza,Lowdin:1955zz}. Consequently, the qubit formulation of the Hamiltonian in finite density becomes
\begin{eqnarray}
H_{\mu}=H-\mu \sum_{n=0}^{N-1}(\sigma_{2n}^{+}\sigma_{2n}^{-}+\sigma_{2n+1}^{+}\sigma_{2n+1}^{-}).
\label{eq:q}
\end{eqnarray}

In this paper, we evaluate explicitly the order parameter for spontaneous chiral symmetry breaking in SU(2) non-Abelian gauge theory at finite temperatures and finite density. The order parameter is identified as the chiral condensate $ \langle \bar{\psi} \psi \rangle$, which can be encoded into the Pauli spin system as follows
\begin{eqnarray}
\langle \bar{\psi} \psi \rangle=\left\langle\sum_{n=0}^{N-1}\left[\left(-1\right)^{n+1}(\sigma_{2n}^{+}\sigma_{2n}^{-}+\sigma_{2n+1}^{+}\sigma_{2n+1}^{-})\right]\right\rangle.
\label{eq:r}
\end{eqnarray}

\subsection{Variational quantum algorithm for chirality imbalance }
In this subsection, we explore the quantum algorithm for calculating chiral condensation at finite temperatures. This approach involves finding the eigenstates of the Hamiltonian and sampling using the Monte Carlo method.

Without loss of generality, we begin with a description of the system's thermodynamic equilibrium state at a given temperature, from the perspective of quantum statistical physics. The Gibbs state \cite{10.5555/1823040} of a certain Hamiltonian is given by
\begin{eqnarray}
\rho(\beta)=\frac{1}{Z(\beta)}e^{-\beta H},\quad Z(\beta)={\rm Tr}\left(e^{-\beta H}\right).
\label{eq:s}
\end{eqnarray}
where $\rho$ is the density matrix and $Z$ is the partition function with $\beta$ the inverse of temperature. Since the Gibbs state corresponds to the minimum of the system's free energy, our motivation to obtain the Gibbs state is to minimize the free energy through variational methods {\cite{Verdon:2019hpy}}. The free energy of a system at finite temperature can be expressed as
\begin{eqnarray}
F(\beta)&=&E(\beta)-TS(\beta)\\\nonumber
&=& \mathrm{Tr}\left[\rho(\beta)H\right] + T \mathrm{Tr}\left[\rho(\beta)\log\rho(\beta)\right],
\label{eq:t}
\end{eqnarray}
where $E(\beta)$ is the average energy, $S(\beta)$ is the von Neumann entropy.

However, if one directly parameterizes the density matrix, the parameters will usually depend on the temperature, and therefore, the optimization must be done at each different temperature. Here, we propose a modified algorithm taking advantage of the Gibbs state being diagonal when expressed by the energy eigenstates $|i\rangle$
\begin{eqnarray}
    \rho=\sum_{i}P_{i}|i\rangle\langle i|,
    \label{eq:u}
\end{eqnarray}
where each state's probability is determined by the corresponding energy $P_{i}=e^{-\beta E_{i}}/\sum_{q}e^{-\beta E_{q}}$. In this expression, only the $P_i$s depend on the temperature, and once the eigenstates are obtained, $P_i$s can be calculated easily. Therefore, we propose a parametrized mixed state as
\begin{eqnarray}
\rho(\theta)=\sum_{i}P_{i}U(\theta)|\varphi_{i}\rangle\langle\varphi_{i}|U(\theta)^{\dagger},
\label{eq:v}
\end{eqnarray}
where $|\varphi_{i}\rangle$ represents initial states from a orthonormal basis. The probability $P_i$ is then given by $P_i = \exp\left[-\beta\langle\varphi_{i}|U(\theta)^{\dagger}HU(\theta)|\varphi_{i}\rangle\right] / Z$. Since $U(\theta)$ is a unitary operator,
the entropy can then be calculated as $S=-\sum_iP_i \log P_i$, which avoids computing the logarithm of the density matrix. The parameters are only related to the energy eigenstates, which means they do not depend on the temperature. Therefore, once they are optimized at one temperature, they are fixed throughout the following calculation. 

The parameterized unitary operator $U(\theta)$ is constructed by the QAOA \cite{Farhi:2014ych,Wiersema:2020ipa}
\begin{eqnarray}
U(\theta)=\prod_{i=1}^{p}\prod_{j=1}^{d}\exp\left(i\theta_{ij}H_{j}\right),
\label{eq:w}
\end{eqnarray}
where the Hamiltonian is decomposed into $H=H_{1}+H_{2}+\cdots+H_{d}$, with $d\geq 2$. Each component $H_{i}$ preserves the same symmetries as the full Hamiltonian $H$ and satisfies the commutation relation $\left[H_{i},H_{j}\right]\neq 0$ with every other component. $p$ is the number of layers in the quantum circuit. Generally, the greater the values of $p$ and $d$, the stronger the representation power of the ansatz.

In practical calculations with a large number of qubits, it is not efficient to traverse all the energy eigenstates. In this paper, we propose to implement a random sampling process as follows
\begin{enumerate}
    \item Randomly select a set of $M$ states $|\varphi_{i}\rangle$ in an orthonormal basis (here we use the computational basis for simplicity), and construct the initial mixed state from the chosen pure states. Then, minimize the free energy with regard to the parameterized mixed state and output the final parameters $\mathbf{\theta}$.
    \item Update the mixed state by choosing another set of $M$ states. Using the previously obtained parameters as initial parameters to perform a variation on the free energy, and subsequently obtain updated parameters.
    \item Repeat step 2 until the parameters do not change considering the designed precision. %changes are smaller than a threshold $\epsilon$.
\end{enumerate}
The advantage of the QAOA is that $U(\theta)$ preserves the same symmetries as $H$. This ensures that all states that are related by conserved symmetries are optimized simultaneously, thus accelerating the optimization.
We have tested that the required number of configurations in each step above is not extensive. Taking 8 qubits as an example, $M=20$ for each step suffices to represent the density matrix of the Gibbs state, which is more efficient than the calculation considering the full 256 eigenstates.

Increasing the number of states used in each step will improve the accuracy of optimization while increasing the computational cost. In addition, using a small set of states can introduce stochasticity and help to avoid falling into local minima during optimization. This strategy is similar to the mini-batch gradient descent method in machine learning, where a subset of data is used in each iteration to balance computational efficiency and optimization performance \cite{8c8eccbbe8a040118afa8f8423da1fe2}.

After obtaining the optimized $U(\theta)$, we perform a Monte Carlo sampling to get the desired density matrix. The steps for Monte Carlo sampling are outlined below
\begin{enumerate}
    \item Initialize a state from the same set of orthonormal basis as used in the optimization steps $|\varphi_{i}\rangle$, calculate the eigenstate $U(\theta)|\varphi_{i}\rangle$ and its energy $E_i$. In practice, we always start from the state $|\varphi_{0}\rangle$ that is mapped to the ground state.
    
    \item Update the input state by randomly flipping one of its qubits and measuring its energy $E_j$ again.
    \item If $E_j < E_i$, accept the new state into the set of states. Otherwise, accept it based on a predetermined acceptance probability $p = e^{-\left(E_j-E_i\right)/T}$. If the new state is rejected, the old state is added to the set again. Notice that in the first step, the energy always increases because we start from the ground state. Therefore, this process starts from the first input state updated from the ground state.
    \item Repeat steps 2 and 3 until a predetermined number of states are generated.
\end{enumerate}
In principle, since the unitary operator $U$ is optimized by a random process, states that are not selected in the optimization steps are not guaranteed to be optimized. However, as discussed above, the symmetry preserving $U$ helps to extend the set of optimized states. Also, since eigenstates with lower energies contribute more to the density matrix, both the optimization and the sampling algorithm will naturally focus on these states; therefore, we do not restrain the states that can be selected in the sampling.

It is easy to understand that the Monte Carlo method will not be efficient for small systems. As shown in the results, for a system with 8 qubits, about 1000 sampled states are used, while the total number of states is just 256. However, when the system becomes large, the total number of states increases exponentially, while the number of sampled states is expected to increase only by a power law.

The above process yields a set of states that effectively reflect the thermodynamic properties of the system, which allows us to calculate the thermal average of an observable $O$ 
\begin{eqnarray}
\langle O \rangle = \frac{1}{N}\sum_{i}\langle\varphi_{i}|U(\theta)^{\dagger}OU(\theta)|\varphi_{i}\rangle.
\label{eq:x}
\end{eqnarray}

Our proposed method has several advantages. Firstly, we do not need any auxiliary qubits. Secondly, the variational calculation, which usually is the most time-consuming, is only done once even if we do our calculation at multiple temperatures. Thirdly, the Monte Carlo sampling enables our method to work even for a large number of qubits. As is shown in the next section, the essential number of sampling states will not grow rapidly with the number of qubits.

\section{Results\label{sec:level3}}
In this section, we present our results of chiral condensate at finite temperature and finite chemical potential within the SU(2) non-Abelian gauge theory. To validate the effectiveness of our quantum algorithm, we compare our simulation results with those obtained from exact diagonalization in 1+1 dimension. The numeral simulation is performed on an open-source package QuSpin\cite{Weinberg:2017igw,Weinberg:2019rfm}. In addition to simulations on classical computers, we also executed our quantum algorithm on IBM’s real quantum hardware \cite{Javadi-Abhari:2024kbf}. 

\begin{figure}[h]
    \includegraphics[width=0.9\linewidth]{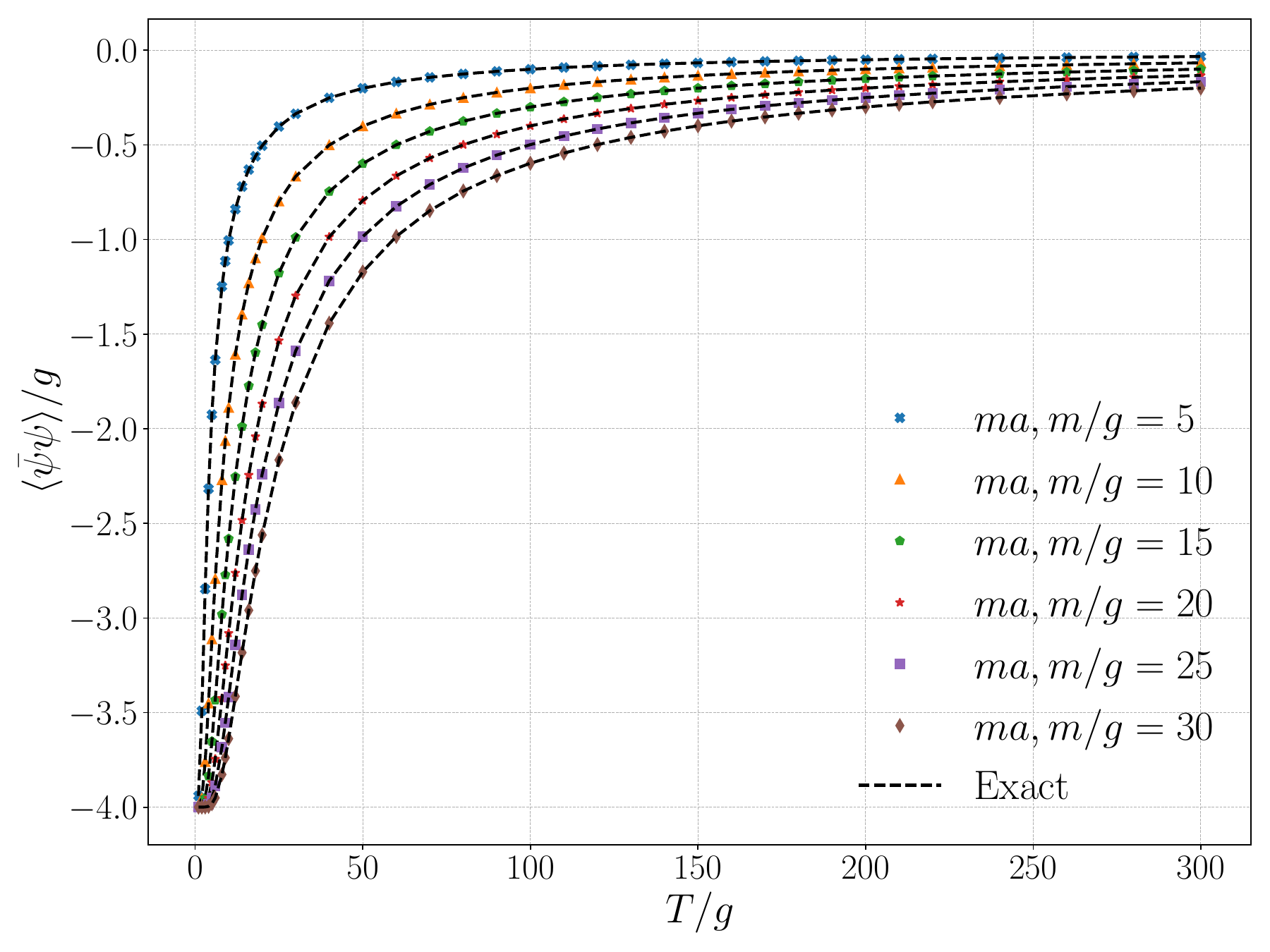}
    \caption{The chiral condensate as a function of temperature $T/g$, in the case of $ma,m/g \in \{5,10,15,20,25,30\}$ with a system size of 8 qubits, where all configurations are considered. }
    \label{fig:a}
\end{figure}

Our simulation is performed with a system size of 8 qubits and a circuit depth of $p=3$. The chiral condensate obtained from the VQA and exact diagonalization is shown in Fig. \ref{fig:a} considering all computational bases. The lines and dots with different colors correspond to different model parameters of $m/g$. The agreement between VQA and exact diagonalization at various $m/g$ validates our proposed quantum algorithm. In particular, at $T\rightarrow 0$, for all values of $m/g$, $\langle \bar{\psi}\psi\rangle /g$ converges to a fixed value of -4. Since we are considering a system of 8 qubits with two spatial points, the maximum expected value of chiral condensate is -4. As the temperature increases, the chiral condensate decreases. However, as the chiral condensate only vanishes at 
$T \to \infty$, there is no phase transition involving the restoration of chiral symmetry. This is consistent with theoretical calculations shown in Refs. \cite{Chelabi:2015cwn,Wetterich:2001dq}.

\begin{figure}[h]
    \includegraphics[width=1\linewidth]{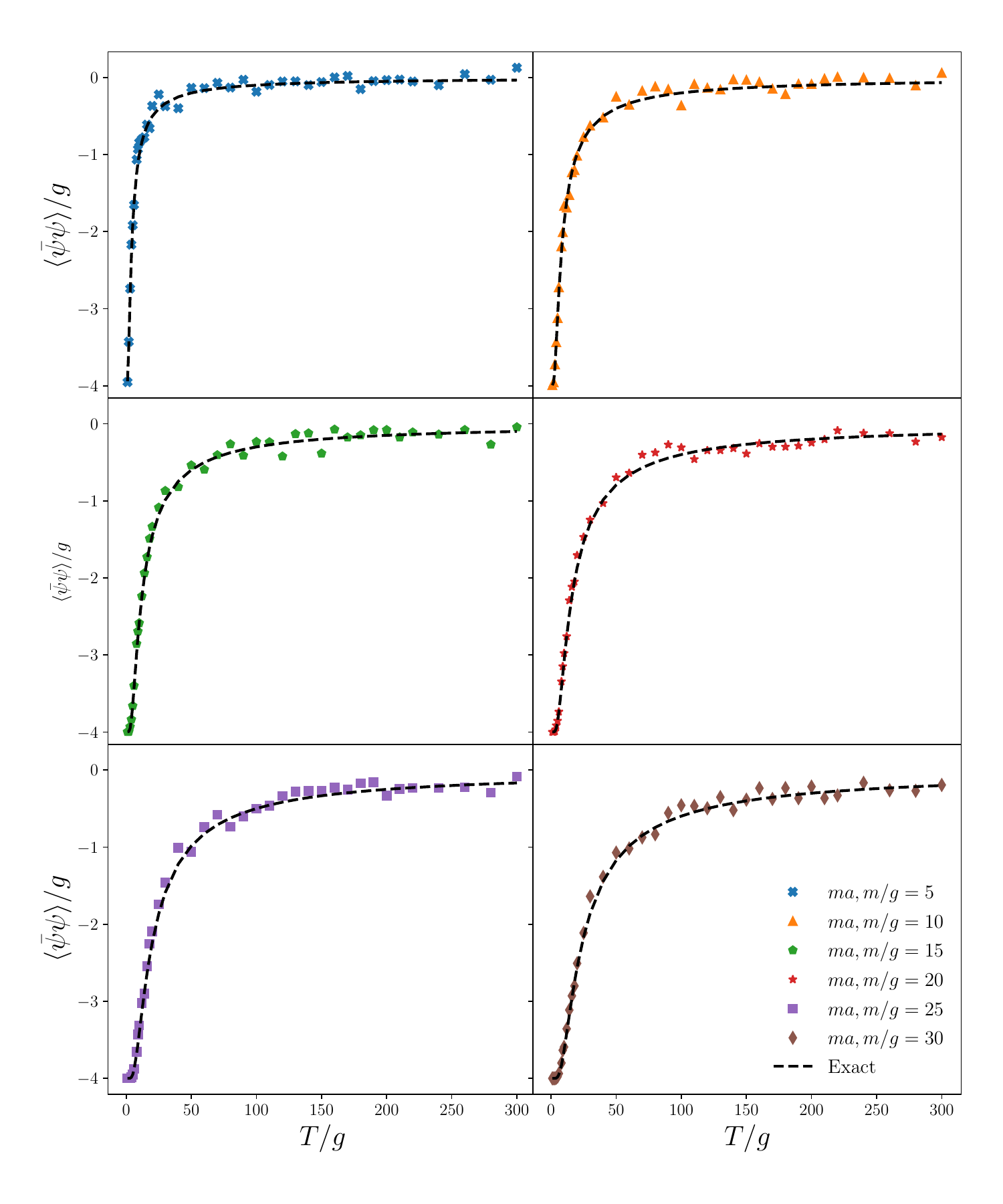}
    \caption{The chiral condensate as a function of temperature $T/g$, in the case of $ma,m/g\in \{5,10,15,20,25,30\}$, with a system size of 8 qubits, which is obtained from MC calculation using 1000 sampled states.}
    \label{fig:b}
\end{figure}

\begin{figure}[h]
    \includegraphics[width=1\linewidth]{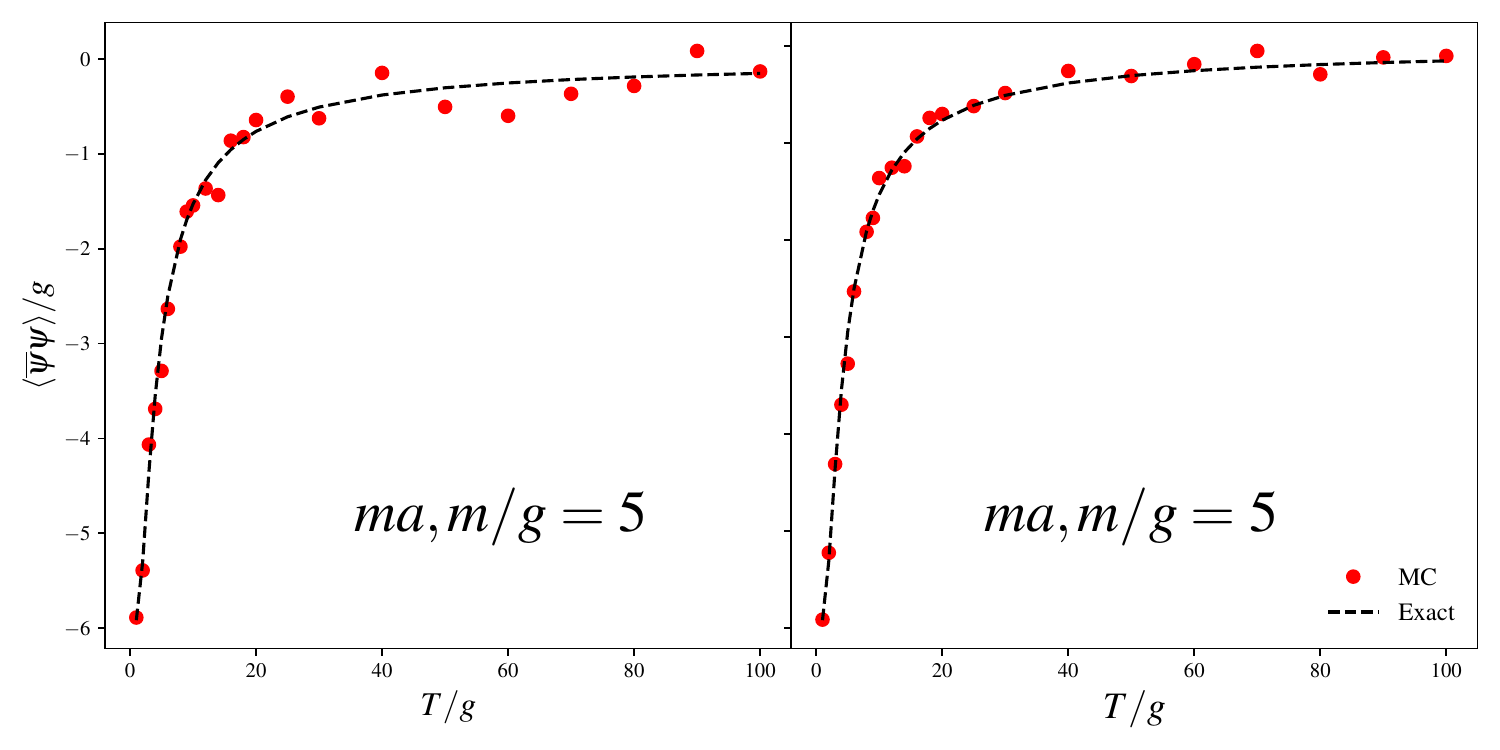}
    \caption{The chiral condensate as a function of temperature $T/g$, in the case of $ma,m/g =5$ with a system size of 12 qubits. The left figure shows the results of sampling 1,000 states, while the right figure presents the results from sampling 2,000 states.}
    \label{fig:c}
\end{figure}

The results from MC calculation are presented in Fig. \ref{fig:b}, in which we sample 1,000 states from a system with 8 qubits. The consistency with the exact diagonalization indicates the capability of the MC sampling method. This may seem inefficient since the system has only 256 computational bases. We emphasize that the sampling does not increase exponentially with the number of qubits, which exhibits great advantage of our approach for large qubit systems. We show such an advantage for a system with 12 qubits in Fig. \ref{fig:c}, where the agreements of samplings for both 1,000 and 2,000 states with exact diagonalization demonstrate the efficiency of our Monte Carlo sampling for a large number of qubits.

The statistical errors can be estimated by standard techniques like bootstrapping. However, one must take into consideration that the samples generated by our Monte Carlo method are autocorrelated, so the effective sample size is reduced. Overall, the errors are consistent with the deviation between numerical and theoretical results. Also, on a state-of-the-art quantum computer, the statistical errors should be less important compared to the errors from quantum circuits. Therefore, the statistical errors are not shown in our results.

\begin{figure}[h]
    \includegraphics[width=0.9\linewidth]{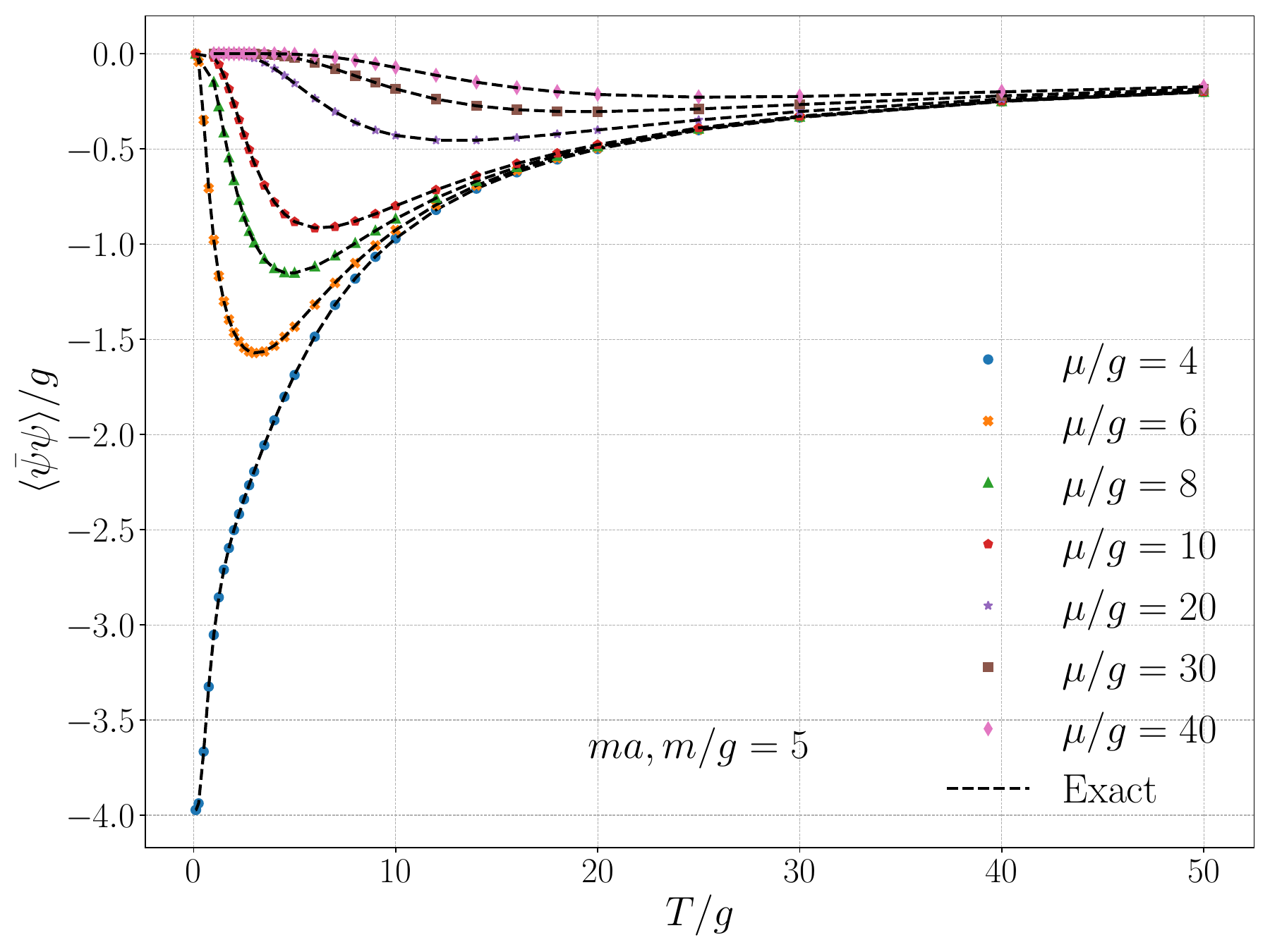}
    \caption{The chiral condensate as a function of temperature $T/g$, in the case of $ma,m/g=5$, with a system size of 8 qubits, at different $\mu/g \in \{4,6,8,10,20,30,40\}$.}
    \label{fig:d}
\end{figure}

\begin{figure}[h]
    \includegraphics[width=0.9\linewidth]{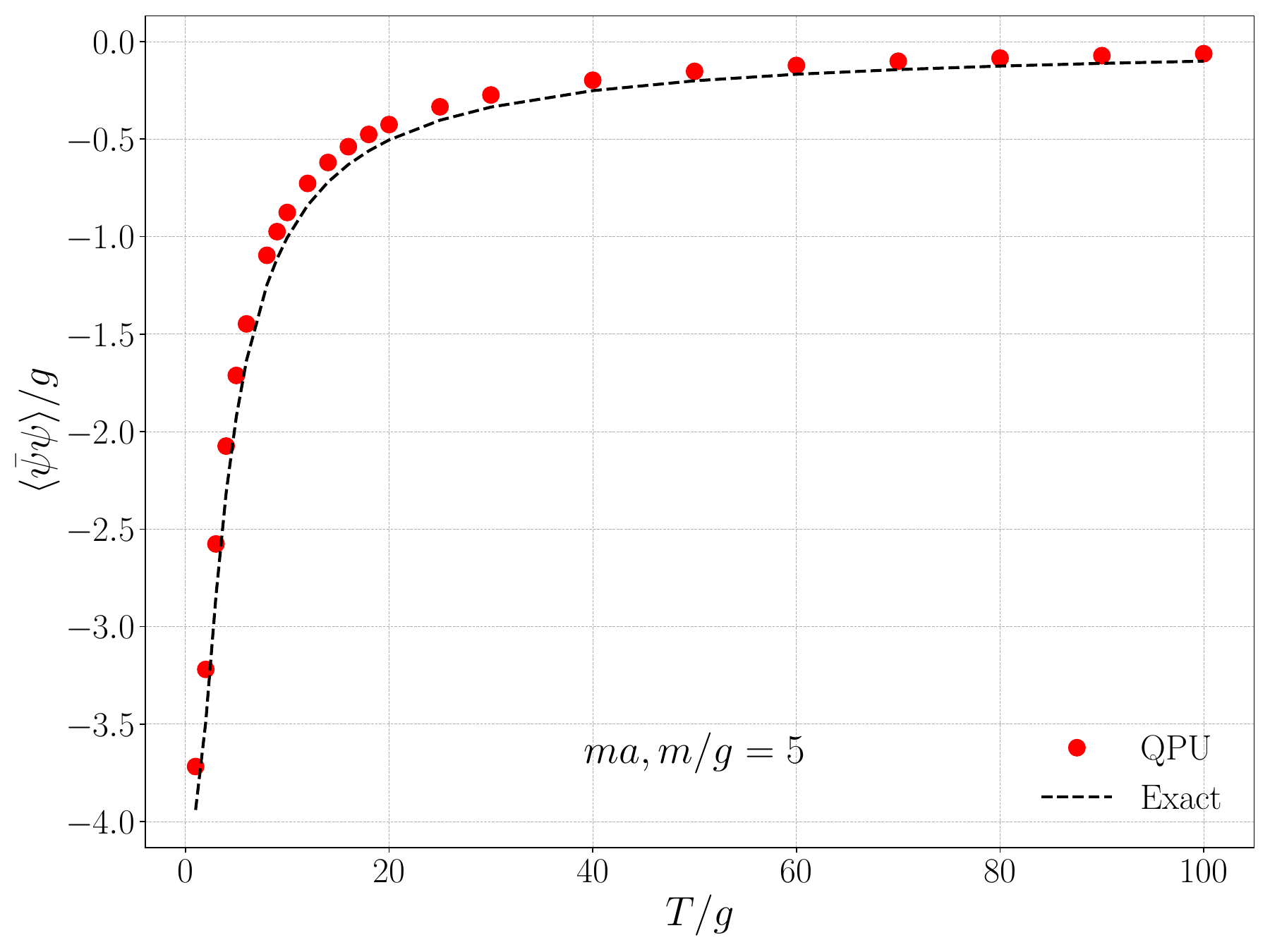}
    \caption{The chiral condensate as a function of temperature $T/g$, in the case of $ma=5$ and $m/g =5$ with a system size of 8 qubits obtained from IBM’s quantum hardware. }
    \label{fig:e}
\end{figure}
\begin{figure}[h]
    \includegraphics[width=0.95\linewidth]{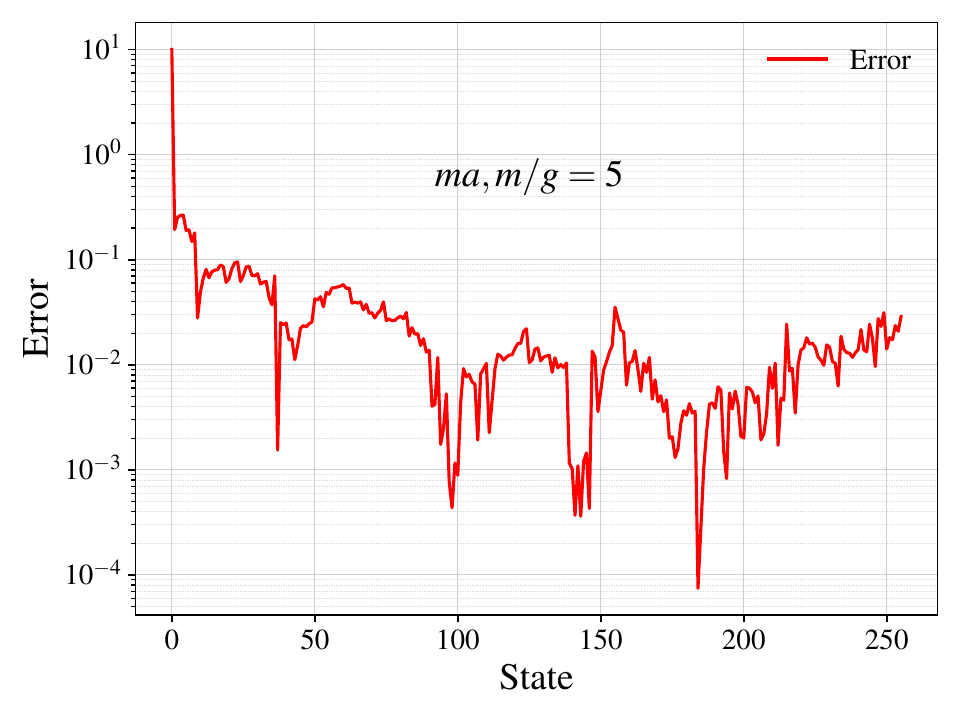}
    \caption{The energy error between the results obtained from IBM’s quantum hardware and the exact diagonalization for the SU(2) non-Abelian gauge theory in the case $ma = 5$ and $m/g = 5$.}
    \label{fig:f}
\end{figure}

We also perform the VQA to calculate the chiral condensate of the SU(2) non-Abelian gauge theory by tuning both temperature and chemical potential. In Fig. \ref{fig:d}, the chiral condensate decreases as the chemical potential increases at the same temperature. At $T \to 0$, the chiral condensate vanishes when the chemical potential is sufficiently large. This corresponds to a change of the ground state. As the temperature increases, the contribution from states with negative expected values of $\langle\bar{\psi}\psi\rangle$ grows, resulting in a non-zero chiral condensate. As the temperature continues to rise, the chiral condensate begins to ``decrease" and eventually vanishes at $T \to \infty$.

To test our proposed algorithm on real quantum hardware, we execute our quantum algorithm on IBM's real quantum hardware, specifically the \textit{ibm\_sherbrooke} device. Owing to the current limitations of quantum resources, we adopted an alternative ansatz as follows
\begin{equation}
U(\theta)=\prod_{l=1}^{p}e^{-i\sum_{i}^{N-1}\xi_{l,i}\sigma^z_{i}\sigma^z_{i+1}}e^{-i\sum_{i}^{N}\alpha_{l,i}\sigma^x_{i}}e^{-i\sum_{i}^{N}\lambda_{l,i}\sigma^z_{i}}.
\end{equation}
This ansatz significantly reduces the circuit depth at the cost of no longer preserving quantum numbers during time evolution. Additionally, due to time constraints, the optimization of parameters is still carried out on a classical computer. After the parameters are optimized, we compute the chiral condensate using all eigenstates of the system on IBM's quantum hardware, with 5000 shots per measurement. The comparison with exact diagonalization is shown in Fig. \ref{fig:e}, and a reasonable good agreement between real quantum computation and exact diagonalization is observed. Furthermore, we compare the eigenenergy obtained from quantum hardware and exact diagonalization. The results for the relative error are displayed in Fig. \ref{fig:f}. In this figure, computational bases are renumbered from 0 to 255 based on the magnitude of their energies. As we can see, the results obtained from IBM's quantum hardware are in good agreement with the ones from exact diagonalization. Notice that the large relative errors for low energy states arise from the addition of a constant to the Hamiltonian, leading to the normalization of the strong coupling ground state energy to zero. The agreement between actual quantum hardware and exact diagonalization indicates the feasibility of using near-term quantum computers for the study of chiral condensation.

\section{Summary and outlook\label{sec:level4}}
In this work, we present the first direct simulation of chiral symmetry breaking using the 1+1D SU(2) non-Abelian gauge theory on a quantum computer. To calculate the chiral condensate as represented by the order parameter for spontaneous chiral symmetry breaking, we propose a variational quantum algorithm to prepare the Gibbs state. We employ the VQA quantum algorithm to calculate the chiral condensate of the SU(2) non-Abelian gauge theory with different $m/g$, temperatures, and chemical potentials. Our simulation results are consistent with those from the exact diagonalization of the discretized Hamiltonian, demonstrating the effectiveness of our quantum algorithm. We emphasize that the employ of Monte Carlo sampling allows the extension of our quantum algorithm to large systems. Our exploration highlights the potential of implementing current or near-term quantum computers to study QCD phase transition, especially at high chemical potential where the Taylor expansion used in classical lattice calculations fails.

\begin{acknowledgments}
This work is supported by the National Natural Science Foundation of China under Grant No. 12035007 and 12475139, Guangdong Major Project of Basic and Applied Basic Research No. 2020B0301030008 and 2023A1515011460.
\end{acknowledgments}

\bibliography{apssamp}

\begin{thebibliography}{69}%
\makeatletter
\providecommand \@ifxundefined [1]{%
 \@ifx{#1\undefined}
}%
\providecommand \@ifnum [1]{%
 \ifnum #1\expandafter \@firstoftwo
 \else \expandafter \@secondoftwo
 \fi
}%
\providecommand \@ifx [1]{%
 \ifx #1\expandafter \@firstoftwo
 \else \expandafter \@secondoftwo
 \fi
}%
\providecommand \natexlab [1]{#1}%
\providecommand \enquote  [1]{``#1''}%
\providecommand \bibnamefont  [1]{#1}%
\providecommand \bibfnamefont [1]{#1}%
\providecommand \citenamefont [1]{#1}%
\providecommand \href@noop [0]{\@secondoftwo}%
\providecommand \href [0]{\begingroup \@sanitize@url \@href}%
\providecommand \@href[1]{\@@startlink{#1}\@@href}%
\providecommand \@@href[1]{\endgroup#1\@@endlink}%
\providecommand \@sanitize@url [0]{\catcode `\\12\catcode `\$12\catcode `\&12\catcode `\#12\catcode `\^12\catcode `\_12\catcode `\%12\relax}%
\providecommand \@@startlink[1]{}%
\providecommand \@@endlink[0]{}%
\providecommand \url  [0]{\begingroup\@sanitize@url \@url }%
\providecommand \@url [1]{\endgroup\@href {#1}{\urlprefix }}%
\providecommand \urlprefix  [0]{URL }%
\providecommand \Eprint [0]{\href }%
\providecommand \doibase [0]{https://doi.org/}%
\providecommand \selectlanguage [0]{\@gobble}%
\providecommand \bibinfo  [0]{\@secondoftwo}%
\providecommand \bibfield  [0]{\@secondoftwo}%
\providecommand \translation [1]{[#1]}%
\providecommand \BibitemOpen [0]{}%
\providecommand \bibitemStop [0]{}%
\providecommand \bibitemNoStop [0]{.\EOS\space}%
\providecommand \EOS [0]{\spacefactor3000\relax}%
\providecommand \BibitemShut  [1]{\csname bibitem#1\endcsname}%
\let\auto@bib@innerbib\@empty
%</preamble>
\bibitem [{\citenamefont {Weinberg}(2005)}]{Weinberg:1995mt}%
  \BibitemOpen
  \bibfield  {author} {\bibinfo {author} {\bibfnamefont {S.}~\bibnamefont {Weinberg}},\ }\href {https://doi.org/10.1017/CBO9781139644167} {\emph {\bibinfo {title} {{The Quantum theory of fields. Vol. 1: Foundations}}}}\ (\bibinfo  {publisher} {Cambridge University Press},\ \bibinfo {year} {2005})\BibitemShut {NoStop}%
\bibitem [{\citenamefont {Gell-Mann}\ \emph {et~al.}(1968)\citenamefont {Gell-Mann}, \citenamefont {Oakes},\ and\ \citenamefont {Renner}}]{Gell-Mann:1968hlm}%
  \BibitemOpen
  \bibfield  {author} {\bibinfo {author} {\bibfnamefont {M.}~\bibnamefont {Gell-Mann}}, \bibinfo {author} {\bibfnamefont {R.~J.}\ \bibnamefont {Oakes}},\ and\ \bibinfo {author} {\bibfnamefont {B.}~\bibnamefont {Renner}},\ }\bibfield  {title} {\bibinfo {title} {{Behavior of current divergences under SU(3) x SU(3)}},\ }\href {https://doi.org/10.1103/PhysRev.175.2195} {\bibfield  {journal} {\bibinfo  {journal} {Phys. Rev.}\ }\textbf {\bibinfo {volume} {175}},\ \bibinfo {pages} {2195} (\bibinfo {year} {1968})}\BibitemShut {NoStop}%
\bibitem [{\citenamefont {Nambu}\ and\ \citenamefont {Jona-Lasinio}(1961{\natexlab{a}})}]{Nambu:1961tp}%
  \BibitemOpen
  \bibfield  {author} {\bibinfo {author} {\bibfnamefont {Y.}~\bibnamefont {Nambu}}\ and\ \bibinfo {author} {\bibfnamefont {G.}~\bibnamefont {Jona-Lasinio}},\ }\bibfield  {title} {\bibinfo {title} {{Dynamical Model of Elementary Particles Based on an Analogy with Superconductivity. 1.}},\ }\href {https://doi.org/10.1103/PhysRev.122.345} {\bibfield  {journal} {\bibinfo  {journal} {Phys. Rev.}\ }\textbf {\bibinfo {volume} {122}},\ \bibinfo {pages} {345} (\bibinfo {year} {1961}{\natexlab{a}})}\BibitemShut {NoStop}%
\bibitem [{\citenamefont {Nambu}\ and\ \citenamefont {Jona-Lasinio}(1961{\natexlab{b}})}]{Nambu:1961fr}%
  \BibitemOpen
  \bibfield  {author} {\bibinfo {author} {\bibfnamefont {Y.}~\bibnamefont {Nambu}}\ and\ \bibinfo {author} {\bibfnamefont {G.}~\bibnamefont {Jona-Lasinio}},\ }\bibfield  {title} {\bibinfo {title} {{Dynamical model of elementary particles based on an analogy with superconductivity. II.}},\ }\href {https://doi.org/10.1103/PhysRev.124.246} {\bibfield  {journal} {\bibinfo  {journal} {Phys. Rev.}\ }\textbf {\bibinfo {volume} {124}},\ \bibinfo {pages} {246} (\bibinfo {year} {1961}{\natexlab{b}})}\BibitemShut {NoStop}%
\bibitem [{\citenamefont {Goldstone}(1961)}]{Goldstone:1961eq}%
  \BibitemOpen
  \bibfield  {author} {\bibinfo {author} {\bibfnamefont {J.}~\bibnamefont {Goldstone}},\ }\bibfield  {title} {\bibinfo {title} {{Field Theories with Superconductor Solutions}},\ }\href {https://doi.org/10.1007/BF02812722} {\bibfield  {journal} {\bibinfo  {journal} {Nuovo Cim.}\ }\textbf {\bibinfo {volume} {19}},\ \bibinfo {pages} {154} (\bibinfo {year} {1961})}\BibitemShut {NoStop}%
\bibitem [{\citenamefont {Shifman}\ \emph {et~al.}(1979)\citenamefont {Shifman}, \citenamefont {Vainshtein},\ and\ \citenamefont {Zakharov}}]{Shifman:1978bx}%
  \BibitemOpen
  \bibfield  {author} {\bibinfo {author} {\bibfnamefont {M.~A.}\ \bibnamefont {Shifman}}, \bibinfo {author} {\bibfnamefont {A.~I.}\ \bibnamefont {Vainshtein}},\ and\ \bibinfo {author} {\bibfnamefont {V.~I.}\ \bibnamefont {Zakharov}},\ }\bibfield  {title} {\bibinfo {title} {{QCD and Resonance Physics. Theoretical Foundations}},\ }\href {https://doi.org/10.1016/0550-3213(79)90022-1} {\bibfield  {journal} {\bibinfo  {journal} {Nucl. Phys. B}\ }\textbf {\bibinfo {volume} {147}},\ \bibinfo {pages} {385} (\bibinfo {year} {1979})}\BibitemShut {NoStop}%
\bibitem [{\citenamefont {Weinberg}(1975)}]{Weinberg:1975ui}%
  \BibitemOpen
  \bibfield  {author} {\bibinfo {author} {\bibfnamefont {S.}~\bibnamefont {Weinberg}},\ }\bibfield  {title} {\bibinfo {title} {{The U(1) Problem}},\ }\href {https://doi.org/10.1103/PhysRevD.11.3583} {\bibfield  {journal} {\bibinfo  {journal} {Phys. Rev. D}\ }\textbf {\bibinfo {volume} {11}},\ \bibinfo {pages} {3583} (\bibinfo {year} {1975})}\BibitemShut {NoStop}%
\bibitem [{\citenamefont {Pisarski}\ and\ \citenamefont {Wilczek}(1984)}]{Pisarski:1983ms}%
  \BibitemOpen
  \bibfield  {author} {\bibinfo {author} {\bibfnamefont {R.~D.}\ \bibnamefont {Pisarski}}\ and\ \bibinfo {author} {\bibfnamefont {F.}~\bibnamefont {Wilczek}},\ }\bibfield  {title} {\bibinfo {title} {{Remarks on the Chiral Phase Transition in Chromodynamics}},\ }\href {https://doi.org/10.1103/PhysRevD.29.338} {\bibfield  {journal} {\bibinfo  {journal} {Phys. Rev. D}\ }\textbf {\bibinfo {volume} {29}},\ \bibinfo {pages} {338} (\bibinfo {year} {1984})}\BibitemShut {NoStop}%
\bibitem [{\citenamefont {Alford}\ \emph {et~al.}(1999)\citenamefont {Alford}, \citenamefont {Rajagopal},\ and\ \citenamefont {Wilczek}}]{Alford:1998mk}%
  \BibitemOpen
  \bibfield  {author} {\bibinfo {author} {\bibfnamefont {M.~G.}\ \bibnamefont {Alford}}, \bibinfo {author} {\bibfnamefont {K.}~\bibnamefont {Rajagopal}},\ and\ \bibinfo {author} {\bibfnamefont {F.}~\bibnamefont {Wilczek}},\ }\bibfield  {title} {\bibinfo {title} {{Color flavor locking and chiral symmetry breaking in high density QCD}},\ }\href {https://doi.org/10.1016/S0550-3213(98)00668-3} {\bibfield  {journal} {\bibinfo  {journal} {Nucl. Phys. B}\ }\textbf {\bibinfo {volume} {537}},\ \bibinfo {pages} {443} (\bibinfo {year} {1999})},\ \Eprint {https://arxiv.org/abs/hep-ph/9804403} {arXiv:hep-ph/9804403} \BibitemShut {NoStop}%
\bibitem [{\citenamefont {Rapp}\ \emph {et~al.}(1998)\citenamefont {Rapp}, \citenamefont {Sch\"afer}, \citenamefont {Shuryak},\ and\ \citenamefont {Velkovsky}}]{Rapp:1997zu}%
  \BibitemOpen
  \bibfield  {author} {\bibinfo {author} {\bibfnamefont {R.}~\bibnamefont {Rapp}}, \bibinfo {author} {\bibfnamefont {T.}~\bibnamefont {Sch\"afer}}, \bibinfo {author} {\bibfnamefont {E.~V.}\ \bibnamefont {Shuryak}},\ and\ \bibinfo {author} {\bibfnamefont {M.}~\bibnamefont {Velkovsky}},\ }\bibfield  {title} {\bibinfo {title} {{Diquark Bose condensates in high density matter and instantons}},\ }\href {https://doi.org/10.1103/PhysRevLett.81.53} {\bibfield  {journal} {\bibinfo  {journal} {Phys. Rev. Lett.}\ }\textbf {\bibinfo {volume} {81}},\ \bibinfo {pages} {53} (\bibinfo {year} {1998})},\ \Eprint {https://arxiv.org/abs/hep-ph/9711396} {arXiv:hep-ph/9711396} \BibitemShut {NoStop}%
\bibitem [{\citenamefont {Alford}\ \emph {et~al.}(2008)\citenamefont {Alford}, \citenamefont {Schmitt}, \citenamefont {Rajagopal},\ and\ \citenamefont {Sch\"afer}}]{Alford:2007xm}%
  \BibitemOpen
  \bibfield  {author} {\bibinfo {author} {\bibfnamefont {M.~G.}\ \bibnamefont {Alford}}, \bibinfo {author} {\bibfnamefont {A.}~\bibnamefont {Schmitt}}, \bibinfo {author} {\bibfnamefont {K.}~\bibnamefont {Rajagopal}},\ and\ \bibinfo {author} {\bibfnamefont {T.}~\bibnamefont {Sch\"afer}},\ }\bibfield  {title} {\bibinfo {title} {{Color superconductivity in dense quark matter}},\ }\href {https://doi.org/10.1103/RevModPhys.80.1455} {\bibfield  {journal} {\bibinfo  {journal} {Rev. Mod. Phys.}\ }\textbf {\bibinfo {volume} {80}},\ \bibinfo {pages} {1455} (\bibinfo {year} {2008})},\ \Eprint {https://arxiv.org/abs/0709.4635} {arXiv:0709.4635 [hep-ph]} \BibitemShut {NoStop}%
\bibitem [{\citenamefont {Rajagopal}\ and\ \citenamefont {Wilczek}(2000)}]{Rajagopal:2000wf}%
  \BibitemOpen
  \bibfield  {author} {\bibinfo {author} {\bibfnamefont {K.}~\bibnamefont {Rajagopal}}\ and\ \bibinfo {author} {\bibfnamefont {F.}~\bibnamefont {Wilczek}},\ }\bibinfo {title} {{The Condensed matter physics of QCD}},\ in\ \href {https://doi.org/10.1142/9789812810458_0043} {\emph {\bibinfo {booktitle} {{At the frontier of particle physics. Handbook of QCD. Vol. 1-3}}}},\ \bibinfo {editor} {edited by\ \bibinfo {editor} {\bibfnamefont {M.}~\bibnamefont {Shifman}}\ and\ \bibinfo {editor} {\bibfnamefont {B.}~\bibnamefont {Ioffe}}}\ (\bibinfo {year} {2000})\ pp.\ \bibinfo {pages} {2061--2151},\ \Eprint {https://arxiv.org/abs/hep-ph/0011333} {arXiv:hep-ph/0011333} \BibitemShut {NoStop}%
\bibitem [{\citenamefont {Stephanov}\ \emph {et~al.}(1998)\citenamefont {Stephanov}, \citenamefont {Rajagopal},\ and\ \citenamefont {Shuryak}}]{Stephanov:1998dy}%
  \BibitemOpen
  \bibfield  {author} {\bibinfo {author} {\bibfnamefont {M.~A.}\ \bibnamefont {Stephanov}}, \bibinfo {author} {\bibfnamefont {K.}~\bibnamefont {Rajagopal}},\ and\ \bibinfo {author} {\bibfnamefont {E.~V.}\ \bibnamefont {Shuryak}},\ }\bibfield  {title} {\bibinfo {title} {{Signatures of the tricritical point in QCD}},\ }\href {https://doi.org/10.1103/PhysRevLett.81.4816} {\bibfield  {journal} {\bibinfo  {journal} {Phys. Rev. Lett.}\ }\textbf {\bibinfo {volume} {81}},\ \bibinfo {pages} {4816} (\bibinfo {year} {1998})},\ \Eprint {https://arxiv.org/abs/hep-ph/9806219} {arXiv:hep-ph/9806219} \BibitemShut {NoStop}%
\bibitem [{\citenamefont {Fukushima}\ and\ \citenamefont {Hatsuda}(2011)}]{Fukushima:2010bq}%
  \BibitemOpen
  \bibfield  {author} {\bibinfo {author} {\bibfnamefont {K.}~\bibnamefont {Fukushima}}\ and\ \bibinfo {author} {\bibfnamefont {T.}~\bibnamefont {Hatsuda}},\ }\bibfield  {title} {\bibinfo {title} {{The phase diagram of dense QCD}},\ }\href {https://doi.org/10.1088/0034-4885/74/1/014001} {\bibfield  {journal} {\bibinfo  {journal} {Rept. Prog. Phys.}\ }\textbf {\bibinfo {volume} {74}},\ \bibinfo {pages} {014001} (\bibinfo {year} {2011})},\ \Eprint {https://arxiv.org/abs/1005.4814} {arXiv:1005.4814 [hep-ph]} \BibitemShut {NoStop}%
\bibitem [{\citenamefont {Gross}\ and\ \citenamefont {Wilczek}(1973)}]{Gross:1973id}%
  \BibitemOpen
  \bibfield  {author} {\bibinfo {author} {\bibfnamefont {D.~J.}\ \bibnamefont {Gross}}\ and\ \bibinfo {author} {\bibfnamefont {F.}~\bibnamefont {Wilczek}},\ }\bibfield  {title} {\bibinfo {title} {{Ultraviolet Behavior of Nonabelian Gauge Theories}},\ }\href {https://doi.org/10.1103/PhysRevLett.30.1343} {\bibfield  {journal} {\bibinfo  {journal} {Phys. Rev. Lett.}\ }\textbf {\bibinfo {volume} {30}},\ \bibinfo {pages} {1343} (\bibinfo {year} {1973})}\BibitemShut {NoStop}%
\bibitem [{\citenamefont {Politzer}(1973)}]{Politzer:1973fx}%
  \BibitemOpen
  \bibfield  {author} {\bibinfo {author} {\bibfnamefont {H.~D.}\ \bibnamefont {Politzer}},\ }\bibfield  {title} {\bibinfo {title} {{Reliable Perturbative Results for Strong Interactions?}},\ }\href {https://doi.org/10.1103/PhysRevLett.30.1346} {\bibfield  {journal} {\bibinfo  {journal} {Phys. Rev. Lett.}\ }\textbf {\bibinfo {volume} {30}},\ \bibinfo {pages} {1346} (\bibinfo {year} {1973})}\BibitemShut {NoStop}%
\bibitem [{\citenamefont {Wilson}(1974)}]{Wilson:1974sk}%
  \BibitemOpen
  \bibfield  {author} {\bibinfo {author} {\bibfnamefont {K.~G.}\ \bibnamefont {Wilson}},\ }\bibfield  {title} {\bibinfo {title} {{Confinement of Quarks}},\ }\href {https://doi.org/10.1103/PhysRevD.10.2445} {\bibfield  {journal} {\bibinfo  {journal} {Phys. Rev. D}\ }\textbf {\bibinfo {volume} {10}},\ \bibinfo {pages} {2445} (\bibinfo {year} {1974})}\BibitemShut {NoStop}%
\bibitem [{\citenamefont {Kogut}(1979)}]{Kogut:1979wt}%
  \BibitemOpen
  \bibfield  {author} {\bibinfo {author} {\bibfnamefont {J.~B.}\ \bibnamefont {Kogut}},\ }\bibfield  {title} {\bibinfo {title} {{An Introduction to Lattice Gauge Theory and Spin Systems}},\ }\href {https://doi.org/10.1103/RevModPhys.51.659} {\bibfield  {journal} {\bibinfo  {journal} {Rev. Mod. Phys.}\ }\textbf {\bibinfo {volume} {51}},\ \bibinfo {pages} {659} (\bibinfo {year} {1979})}\BibitemShut {NoStop}%
\bibitem [{\citenamefont {Aoki}\ \emph {et~al.}(2017)\citenamefont {Aoki} \emph {et~al.}}]{Aoki:2016frl}%
  \BibitemOpen
  \bibfield  {author} {\bibinfo {author} {\bibfnamefont {S.}~\bibnamefont {Aoki}} \emph {et~al.},\ }\bibfield  {title} {\bibinfo {title} {{Review of lattice results concerning low-energy particle physics}},\ }\href {https://doi.org/10.1140/epjc/s10052-016-4509-7} {\bibfield  {journal} {\bibinfo  {journal} {Eur. Phys. J. C}\ }\textbf {\bibinfo {volume} {77}},\ \bibinfo {pages} {112} (\bibinfo {year} {2017})},\ \Eprint {https://arxiv.org/abs/1607.00299} {arXiv:1607.00299 [hep-lat]} \BibitemShut {NoStop}%
\bibitem [{\citenamefont {Durr}\ \emph {et~al.}(2008)\citenamefont {Durr} \emph {et~al.}}]{BMW:2008jgk}%
  \BibitemOpen
  \bibfield  {author} {\bibinfo {author} {\bibfnamefont {S.}~\bibnamefont {Durr}} \emph {et~al.} (\bibinfo {collaboration} {BMW}),\ }\bibfield  {title} {\bibinfo {title} {{Ab-Initio Determination of Light Hadron Masses}},\ }\href {https://doi.org/10.1126/science.1163233} {\bibfield  {journal} {\bibinfo  {journal} {Science}\ }\textbf {\bibinfo {volume} {322}},\ \bibinfo {pages} {1224} (\bibinfo {year} {2008})},\ \Eprint {https://arxiv.org/abs/0906.3599} {arXiv:0906.3599 [hep-lat]} \BibitemShut {NoStop}%
\bibitem [{\citenamefont {Troyer}\ and\ \citenamefont {Wiese}(2005)}]{Troyer:2004ge}%
  \BibitemOpen
  \bibfield  {author} {\bibinfo {author} {\bibfnamefont {M.}~\bibnamefont {Troyer}}\ and\ \bibinfo {author} {\bibfnamefont {U.-J.}\ \bibnamefont {Wiese}},\ }\bibfield  {title} {\bibinfo {title} {{Computational complexity and fundamental limitations to fermionic quantum Monte Carlo simulations}},\ }\href {https://doi.org/10.1103/PhysRevLett.94.170201} {\bibfield  {journal} {\bibinfo  {journal} {Phys. Rev. Lett.}\ }\textbf {\bibinfo {volume} {94}},\ \bibinfo {pages} {170201} (\bibinfo {year} {2005})},\ \Eprint {https://arxiv.org/abs/cond-mat/0408370} {arXiv:cond-mat/0408370} \BibitemShut {NoStop}%
\bibitem [{\citenamefont {de~Forcrand}\ and\ \citenamefont {Philipsen}(2002)}]{deForcrand:2002hgr}%
  \BibitemOpen
  \bibfield  {author} {\bibinfo {author} {\bibfnamefont {P.}~\bibnamefont {de~Forcrand}}\ and\ \bibinfo {author} {\bibfnamefont {O.}~\bibnamefont {Philipsen}},\ }\bibfield  {title} {\bibinfo {title} {{The QCD phase diagram for small densities from imaginary chemical potential}},\ }\href {https://doi.org/10.1016/S0550-3213(02)00626-0} {\bibfield  {journal} {\bibinfo  {journal} {Nucl. Phys. B}\ }\textbf {\bibinfo {volume} {642}},\ \bibinfo {pages} {290} (\bibinfo {year} {2002})},\ \Eprint {https://arxiv.org/abs/hep-lat/0205016} {arXiv:hep-lat/0205016} \BibitemShut {NoStop}%
\bibitem [{\citenamefont {Bauer}\ \emph {et~al.}(2023)\citenamefont {Bauer} \emph {et~al.}}]{Bauer:2022hpo}%
  \BibitemOpen
  \bibfield  {author} {\bibinfo {author} {\bibfnamefont {C.~W.}\ \bibnamefont {Bauer}} \emph {et~al.},\ }\bibfield  {title} {\bibinfo {title} {{Quantum Simulation for High-Energy Physics}},\ }\href {https://doi.org/10.1103/PRXQuantum.4.027001} {\bibfield  {journal} {\bibinfo  {journal} {PRX Quantum}\ }\textbf {\bibinfo {volume} {4}},\ \bibinfo {pages} {027001} (\bibinfo {year} {2023})},\ \Eprint {https://arxiv.org/abs/2204.03381} {arXiv:2204.03381 [quant-ph]} \BibitemShut {NoStop}%
\bibitem [{\citenamefont {Di~Meglio}\ \emph {et~al.}(2024)\citenamefont {Di~Meglio} \emph {et~al.}}]{DiMeglio:2023nsa}%
  \BibitemOpen
  \bibfield  {author} {\bibinfo {author} {\bibfnamefont {A.}~\bibnamefont {Di~Meglio}} \emph {et~al.},\ }\bibfield  {title} {\bibinfo {title} {{Quantum Computing for High-Energy Physics: State of the Art and Challenges}},\ }\href {https://doi.org/10.1103/PRXQuantum.5.037001} {\bibfield  {journal} {\bibinfo  {journal} {PRX Quantum}\ }\textbf {\bibinfo {volume} {5}},\ \bibinfo {pages} {037001} (\bibinfo {year} {2024})},\ \Eprint {https://arxiv.org/abs/2307.03236} {arXiv:2307.03236 [quant-ph]} \BibitemShut {NoStop}%
\bibitem [{\citenamefont {Zhang}\ \emph {et~al.}(2021)\citenamefont {Zhang}, \citenamefont {Xing}, \citenamefont {Yan}, \citenamefont {Wang},\ and\ \citenamefont {Zhu}}]{Zhang:2020uqo}%
  \BibitemOpen
  \bibfield  {author} {\bibinfo {author} {\bibfnamefont {D.-B.}\ \bibnamefont {Zhang}}, \bibinfo {author} {\bibfnamefont {H.}~\bibnamefont {Xing}}, \bibinfo {author} {\bibfnamefont {H.}~\bibnamefont {Yan}}, \bibinfo {author} {\bibfnamefont {E.}~\bibnamefont {Wang}},\ and\ \bibinfo {author} {\bibfnamefont {S.-L.}\ \bibnamefont {Zhu}},\ }\bibfield  {title} {\bibinfo {title} {{Selected topics of quantum computing for nuclear physics}},\ }\href {https://doi.org/10.1088/1674-1056/abd761} {\bibfield  {journal} {\bibinfo  {journal} {Chin. Phys. B}\ }\textbf {\bibinfo {volume} {30}},\ \bibinfo {pages} {020306} (\bibinfo {year} {2021})},\ \Eprint {https://arxiv.org/abs/2011.01431} {arXiv:2011.01431 [quant-ph]} \BibitemShut {NoStop}%
\bibitem [{\citenamefont {Fang}\ \emph {et~al.}(2024)\citenamefont {Fang}, \citenamefont {Gao}, \citenamefont {Li}, \citenamefont {Shu}, \citenamefont {Wu}, \citenamefont {Xing}, \citenamefont {Xu}, \citenamefont {Xu},\ and\ \citenamefont {Zhou}}]{Fang:2024ple}%
  \BibitemOpen
  \bibfield  {author} {\bibinfo {author} {\bibfnamefont {Y.}~\bibnamefont {Fang}}, \bibinfo {author} {\bibfnamefont {C.}~\bibnamefont {Gao}}, \bibinfo {author} {\bibfnamefont {Y.-Y.}\ \bibnamefont {Li}}, \bibinfo {author} {\bibfnamefont {J.}~\bibnamefont {Shu}}, \bibinfo {author} {\bibfnamefont {Y.}~\bibnamefont {Wu}}, \bibinfo {author} {\bibfnamefont {H.}~\bibnamefont {Xing}}, \bibinfo {author} {\bibfnamefont {B.}~\bibnamefont {Xu}}, \bibinfo {author} {\bibfnamefont {L.}~\bibnamefont {Xu}},\ and\ \bibinfo {author} {\bibfnamefont {C.}~\bibnamefont {Zhou}},\ }\bibfield  {title} {\bibinfo {title} {{Quantum Frontiers in High Energy Physics}},\ }\href@noop {} {\  (\bibinfo {year} {2024})},\ \Eprint {https://arxiv.org/abs/2411.11294} {arXiv:2411.11294 [hep-ph]} \BibitemShut {NoStop}%
\bibitem [{\citenamefont {Li}\ \emph {et~al.}(2022)\citenamefont {Li}, \citenamefont {Guo}, \citenamefont {Lai}, \citenamefont {Liu}, \citenamefont {Wang}, \citenamefont {Xing}, \citenamefont {Zhang},\ and\ \citenamefont {Zhu}}]{Li:2021kcs}%
  \BibitemOpen
  \bibfield  {author} {\bibinfo {author} {\bibfnamefont {T.}~\bibnamefont {Li}}, \bibinfo {author} {\bibfnamefont {X.}~\bibnamefont {Guo}}, \bibinfo {author} {\bibfnamefont {W.~K.}\ \bibnamefont {Lai}}, \bibinfo {author} {\bibfnamefont {X.}~\bibnamefont {Liu}}, \bibinfo {author} {\bibfnamefont {E.}~\bibnamefont {Wang}}, \bibinfo {author} {\bibfnamefont {H.}~\bibnamefont {Xing}}, \bibinfo {author} {\bibfnamefont {D.-B.}\ \bibnamefont {Zhang}},\ and\ \bibinfo {author} {\bibfnamefont {S.-L.}\ \bibnamefont {Zhu}} (\bibinfo {collaboration} {QuNu}),\ }\bibfield  {title} {\bibinfo {title} {{Partonic collinear structure by quantum computing}},\ }\href {https://doi.org/10.1103/PhysRevD.105.L111502} {\bibfield  {journal} {\bibinfo  {journal} {Phys. Rev. D}\ }\textbf {\bibinfo {volume} {105}},\ \bibinfo {pages} {L111502} (\bibinfo {year} {2022})},\ \Eprint {https://arxiv.org/abs/2106.03865} {arXiv:2106.03865 [hep-ph]} \BibitemShut {NoStop}%
\bibitem [{\citenamefont {Li}\ \emph {et~al.}(2023)\citenamefont {Li}, \citenamefont {Guo}, \citenamefont {Lai}, \citenamefont {Liu}, \citenamefont {Wang}, \citenamefont {Xing}, \citenamefont {Zhang},\ and\ \citenamefont {Zhu}}]{Li:2022lyt}%
  \BibitemOpen
  \bibfield  {author} {\bibinfo {author} {\bibfnamefont {T.}~\bibnamefont {Li}}, \bibinfo {author} {\bibfnamefont {X.}~\bibnamefont {Guo}}, \bibinfo {author} {\bibfnamefont {W.~K.}\ \bibnamefont {Lai}}, \bibinfo {author} {\bibfnamefont {X.}~\bibnamefont {Liu}}, \bibinfo {author} {\bibfnamefont {E.}~\bibnamefont {Wang}}, \bibinfo {author} {\bibfnamefont {H.}~\bibnamefont {Xing}}, \bibinfo {author} {\bibfnamefont {D.-B.}\ \bibnamefont {Zhang}},\ and\ \bibinfo {author} {\bibfnamefont {S.-L.}\ \bibnamefont {Zhu}} (\bibinfo {collaboration} {QuNu}),\ }\bibfield  {title} {\bibinfo {title} {{Exploring light-cone distribution amplitudes from quantum computing}},\ }\href {https://doi.org/10.1007/s11433-023-2120-1} {\bibfield  {journal} {\bibinfo  {journal} {Sci. China Phys. Mech. Astron.}\ }\textbf {\bibinfo {volume} {66}},\ \bibinfo {pages} {281011} (\bibinfo {year} {2023})},\ \Eprint {https://arxiv.org/abs/2207.13258} {arXiv:2207.13258 [hep-ph]} \BibitemShut {NoStop}%
\bibitem [{\citenamefont {Li}\ \emph {et~al.}(2024{\natexlab{a}})\citenamefont {Li}, \citenamefont {Lai}, \citenamefont {Wang},\ and\ \citenamefont {Xing}}]{Li:2023kex}%
  \BibitemOpen
  \bibfield  {author} {\bibinfo {author} {\bibfnamefont {T.}~\bibnamefont {Li}}, \bibinfo {author} {\bibfnamefont {W.~K.}\ \bibnamefont {Lai}}, \bibinfo {author} {\bibfnamefont {E.}~\bibnamefont {Wang}},\ and\ \bibinfo {author} {\bibfnamefont {H.}~\bibnamefont {Xing}} (\bibinfo {collaboration} {QuNu}),\ }\bibfield  {title} {\bibinfo {title} {{Scattering amplitude from quantum computing with reduction formula}},\ }\href {https://doi.org/10.1103/PhysRevD.109.036025} {\bibfield  {journal} {\bibinfo  {journal} {Phys. Rev. D}\ }\textbf {\bibinfo {volume} {109}},\ \bibinfo {pages} {036025} (\bibinfo {year} {2024}{\natexlab{a}})},\ \Eprint {https://arxiv.org/abs/2301.04179} {arXiv:2301.04179 [hep-ph]} \BibitemShut {NoStop}%
\bibitem [{\citenamefont {Lamm}\ \emph {et~al.}(2020)\citenamefont {Lamm}, \citenamefont {Lawrence},\ and\ \citenamefont {Yamauchi}}]{Lamm:2019uyc}%
  \BibitemOpen
  \bibfield  {author} {\bibinfo {author} {\bibfnamefont {H.}~\bibnamefont {Lamm}}, \bibinfo {author} {\bibfnamefont {S.}~\bibnamefont {Lawrence}},\ and\ \bibinfo {author} {\bibfnamefont {Y.}~\bibnamefont {Yamauchi}} (\bibinfo {collaboration} {NuQS}),\ }\bibfield  {title} {\bibinfo {title} {{Parton physics on a quantum computer}},\ }\href {https://doi.org/10.1103/PhysRevResearch.2.013272} {\bibfield  {journal} {\bibinfo  {journal} {Phys. Rev. Res.}\ }\textbf {\bibinfo {volume} {2}},\ \bibinfo {pages} {013272} (\bibinfo {year} {2020})},\ \Eprint {https://arxiv.org/abs/1908.10439} {arXiv:1908.10439 [hep-lat]} \BibitemShut {NoStop}%
\bibitem [{\citenamefont {Li}\ \emph {et~al.}(2024{\natexlab{b}})\citenamefont {Li}, \citenamefont {Xing},\ and\ \citenamefont {Zhang}}]{Li:2024nod}%
  \BibitemOpen
  \bibfield  {author} {\bibinfo {author} {\bibfnamefont {T.}~\bibnamefont {Li}}, \bibinfo {author} {\bibfnamefont {H.}~\bibnamefont {Xing}},\ and\ \bibinfo {author} {\bibfnamefont {D.-B.}\ \bibnamefont {Zhang}},\ }\bibfield  {title} {\bibinfo {title} {{Simulating Parton Fragmentation on Quantum Computers}},\ }\href@noop {} {\  (\bibinfo {year} {2024}{\natexlab{b}})},\ \Eprint {https://arxiv.org/abs/2406.05683} {arXiv:2406.05683 [hep-ph]} \BibitemShut {NoStop}%
\bibitem [{\citenamefont {Grieninger}\ and\ \citenamefont {Zahed}(2024)}]{Grieninger:2024axp}%
  \BibitemOpen
  \bibfield  {author} {\bibinfo {author} {\bibfnamefont {S.}~\bibnamefont {Grieninger}}\ and\ \bibinfo {author} {\bibfnamefont {I.}~\bibnamefont {Zahed}},\ }\bibfield  {title} {\bibinfo {title} {{Quasi-fragmentation functions in the massive Schwinger model}},\ }\href@noop {} {\  (\bibinfo {year} {2024})},\ \Eprint {https://arxiv.org/abs/2406.01891} {arXiv:2406.01891 [hep-ph]} \BibitemShut {NoStop}%
\bibitem [{\citenamefont {Czajka}\ \emph {et~al.}(2022)\citenamefont {Czajka}, \citenamefont {Kang}, \citenamefont {Ma},\ and\ \citenamefont {Zhao}}]{Czajka:2021yll}%
  \BibitemOpen
  \bibfield  {author} {\bibinfo {author} {\bibfnamefont {A.~M.}\ \bibnamefont {Czajka}}, \bibinfo {author} {\bibfnamefont {Z.-B.}\ \bibnamefont {Kang}}, \bibinfo {author} {\bibfnamefont {H.}~\bibnamefont {Ma}},\ and\ \bibinfo {author} {\bibfnamefont {F.}~\bibnamefont {Zhao}},\ }\bibfield  {title} {\bibinfo {title} {{Quantum simulation of chiral phase transitions}},\ }\href {https://doi.org/10.1007/JHEP08(2022)209} {\bibfield  {journal} {\bibinfo  {journal} {JHEP}\ }\textbf {\bibinfo {volume} {08}},\ \bibinfo {pages} {209}},\ \Eprint {https://arxiv.org/abs/2112.03944} {arXiv:2112.03944 [hep-ph]} \BibitemShut {NoStop}%
\bibitem [{\citenamefont {de~Jong}\ \emph {et~al.}(2022)\citenamefont {de~Jong}, \citenamefont {Lee}, \citenamefont {Mulligan}, \citenamefont {P\l{}osko\'n}, \citenamefont {Ringer},\ and\ \citenamefont {Yao}}]{deJong:2021wsd}%
  \BibitemOpen
  \bibfield  {author} {\bibinfo {author} {\bibfnamefont {W.~A.}\ \bibnamefont {de~Jong}}, \bibinfo {author} {\bibfnamefont {K.}~\bibnamefont {Lee}}, \bibinfo {author} {\bibfnamefont {J.}~\bibnamefont {Mulligan}}, \bibinfo {author} {\bibfnamefont {M.}~\bibnamefont {P\l{}osko\'n}}, \bibinfo {author} {\bibfnamefont {F.}~\bibnamefont {Ringer}},\ and\ \bibinfo {author} {\bibfnamefont {X.}~\bibnamefont {Yao}},\ }\bibfield  {title} {\bibinfo {title} {{Quantum simulation of nonequilibrium dynamics and thermalization in the Schwinger model}},\ }\href {https://doi.org/10.1103/PhysRevD.106.054508} {\bibfield  {journal} {\bibinfo  {journal} {Phys. Rev. D}\ }\textbf {\bibinfo {volume} {106}},\ \bibinfo {pages} {054508} (\bibinfo {year} {2022})},\ \Eprint {https://arxiv.org/abs/2106.08394} {arXiv:2106.08394 [quant-ph]} \BibitemShut {NoStop}%
\bibitem [{\citenamefont {Yang}\ \emph {et~al.}(2020)\citenamefont {Yang}, \citenamefont {Sun}, \citenamefont {Ott}, \citenamefont {Wang}, \citenamefont {Zache}, \citenamefont {Halimeh}, \citenamefont {Yuan}, \citenamefont {Hauke},\ and\ \citenamefont {Pan}}]{Yang:2020yer}%
  \BibitemOpen
  \bibfield  {author} {\bibinfo {author} {\bibfnamefont {B.}~\bibnamefont {Yang}}, \bibinfo {author} {\bibfnamefont {H.}~\bibnamefont {Sun}}, \bibinfo {author} {\bibfnamefont {R.}~\bibnamefont {Ott}}, \bibinfo {author} {\bibfnamefont {H.-Y.}\ \bibnamefont {Wang}}, \bibinfo {author} {\bibfnamefont {T.~V.}\ \bibnamefont {Zache}}, \bibinfo {author} {\bibfnamefont {J.~C.}\ \bibnamefont {Halimeh}}, \bibinfo {author} {\bibfnamefont {Z.-S.}\ \bibnamefont {Yuan}}, \bibinfo {author} {\bibfnamefont {P.}~\bibnamefont {Hauke}},\ and\ \bibinfo {author} {\bibfnamefont {J.-W.}\ \bibnamefont {Pan}},\ }\bibfield  {title} {\bibinfo {title} {{Observation of gauge invariance in a 71-site Bose\textendash{}Hubbard quantum simulator}},\ }\href {https://doi.org/10.1038/s41586-020-2910-8} {\bibfield  {journal} {\bibinfo  {journal} {Nature}\ }\textbf {\bibinfo {volume} {587}},\ \bibinfo {pages} {392} (\bibinfo {year} {2020})},\ \Eprint {https://arxiv.org/abs/2003.08945} {arXiv:2003.08945 [cond-mat.quant-gas]} \BibitemShut {NoStop}%
\bibitem [{\citenamefont {Zhou}\ \emph {et~al.}(2022)\citenamefont {Zhou}, \citenamefont {Su}, \citenamefont {Halimeh}, \citenamefont {Ott}, \citenamefont {Sun}, \citenamefont {Hauke}, \citenamefont {Yang}, \citenamefont {Yuan}, \citenamefont {Berges},\ and\ \citenamefont {Pan}}]{Zhou:2021kdl}%
  \BibitemOpen
  \bibfield  {author} {\bibinfo {author} {\bibfnamefont {Z.-Y.}\ \bibnamefont {Zhou}}, \bibinfo {author} {\bibfnamefont {G.-X.}\ \bibnamefont {Su}}, \bibinfo {author} {\bibfnamefont {J.~C.}\ \bibnamefont {Halimeh}}, \bibinfo {author} {\bibfnamefont {R.}~\bibnamefont {Ott}}, \bibinfo {author} {\bibfnamefont {H.}~\bibnamefont {Sun}}, \bibinfo {author} {\bibfnamefont {P.}~\bibnamefont {Hauke}}, \bibinfo {author} {\bibfnamefont {B.}~\bibnamefont {Yang}}, \bibinfo {author} {\bibfnamefont {Z.-S.}\ \bibnamefont {Yuan}}, \bibinfo {author} {\bibfnamefont {J.}~\bibnamefont {Berges}},\ and\ \bibinfo {author} {\bibfnamefont {J.-W.}\ \bibnamefont {Pan}},\ }\bibfield  {title} {\bibinfo {title} {{Thermalization dynamics of a gauge theory on a quantum simulator}},\ }\href {https://doi.org/10.1126/science.abl6277} {\bibfield  {journal} {\bibinfo  {journal} {Science}\ }\textbf {\bibinfo {volume} {377}},\ \bibinfo {pages} {abl6277} (\bibinfo {year} {2022})},\ \Eprint {https://arxiv.org/abs/2107.13563} {arXiv:2107.13563
  [cond-mat.quant-gas]} \BibitemShut {NoStop}%
\bibitem [{\citenamefont {Kokail}\ \emph {et~al.}(2019)\citenamefont {Kokail} \emph {et~al.}}]{Kokail:2018eiw}%
  \BibitemOpen
  \bibfield  {author} {\bibinfo {author} {\bibfnamefont {C.}~\bibnamefont {Kokail}} \emph {et~al.},\ }\bibfield  {title} {\bibinfo {title} {{Self-verifying variational quantum simulation of lattice models}},\ }\href {https://doi.org/10.1038/s41586-019-1177-4} {\bibfield  {journal} {\bibinfo  {journal} {Nature}\ }\textbf {\bibinfo {volume} {569}},\ \bibinfo {pages} {355} (\bibinfo {year} {2019})},\ \Eprint {https://arxiv.org/abs/1810.03421} {arXiv:1810.03421 [quant-ph]} \BibitemShut {NoStop}%
\bibitem [{\citenamefont {Xie}\ \emph {et~al.}(2022)\citenamefont {Xie}, \citenamefont {Guo}, \citenamefont {Xing}, \citenamefont {Xue}, \citenamefont {Zhang},\ and\ \citenamefont {Zhu}}]{Xie:2022jgj}%
  \BibitemOpen
  \bibfield  {author} {\bibinfo {author} {\bibfnamefont {X.-D.}\ \bibnamefont {Xie}}, \bibinfo {author} {\bibfnamefont {X.}~\bibnamefont {Guo}}, \bibinfo {author} {\bibfnamefont {H.}~\bibnamefont {Xing}}, \bibinfo {author} {\bibfnamefont {Z.-Y.}\ \bibnamefont {Xue}}, \bibinfo {author} {\bibfnamefont {D.-B.}\ \bibnamefont {Zhang}},\ and\ \bibinfo {author} {\bibfnamefont {S.-L.}\ \bibnamefont {Zhu}} (\bibinfo {collaboration} {QuNu}),\ }\bibfield  {title} {\bibinfo {title} {{Variational thermal quantum simulation of the lattice Schwinger model}},\ }\href {https://doi.org/10.1103/PhysRevD.106.054509} {\bibfield  {journal} {\bibinfo  {journal} {Phys. Rev. D}\ }\textbf {\bibinfo {volume} {106}},\ \bibinfo {pages} {054509} (\bibinfo {year} {2022})},\ \Eprint {https://arxiv.org/abs/2205.12767} {arXiv:2205.12767 [quant-ph]} \BibitemShut {NoStop}%
\bibitem [{\citenamefont {Davoudi}\ \emph {et~al.}(2023)\citenamefont {Davoudi}, \citenamefont {Mueller},\ and\ \citenamefont {Powers}}]{Davoudi:2022uzo}%
  \BibitemOpen
  \bibfield  {author} {\bibinfo {author} {\bibfnamefont {Z.}~\bibnamefont {Davoudi}}, \bibinfo {author} {\bibfnamefont {N.}~\bibnamefont {Mueller}},\ and\ \bibinfo {author} {\bibfnamefont {C.}~\bibnamefont {Powers}},\ }\bibfield  {title} {\bibinfo {title} {{Towards Quantum Computing Phase Diagrams of Gauge Theories with Thermal Pure Quantum States}},\ }\href {https://doi.org/10.1103/PhysRevLett.131.081901} {\bibfield  {journal} {\bibinfo  {journal} {Phys. Rev. Lett.}\ }\textbf {\bibinfo {volume} {131}},\ \bibinfo {pages} {081901} (\bibinfo {year} {2023})},\ \Eprint {https://arxiv.org/abs/2208.13112} {arXiv:2208.13112 [hep-lat]} \BibitemShut {NoStop}%
\bibitem [{\citenamefont {Araz}\ \emph {et~al.}(2024)\citenamefont {Araz}, \citenamefont {Jha}, \citenamefont {Ringer},\ and\ \citenamefont {Sambasivam}}]{Araz:2024xkw}%
  \BibitemOpen
  \bibfield  {author} {\bibinfo {author} {\bibfnamefont {J.~Y.}\ \bibnamefont {Araz}}, \bibinfo {author} {\bibfnamefont {R.~G.}\ \bibnamefont {Jha}}, \bibinfo {author} {\bibfnamefont {F.}~\bibnamefont {Ringer}},\ and\ \bibinfo {author} {\bibfnamefont {B.}~\bibnamefont {Sambasivam}},\ }\bibfield  {title} {\bibinfo {title} {{Thermal state preparation of the SYK model using a variational quantum algorithm}},\ }\href@noop {} {\  (\bibinfo {year} {2024})},\ \Eprint {https://arxiv.org/abs/2406.15545} {arXiv:2406.15545 [quant-ph]} \BibitemShut {NoStop}%
\bibitem [{\citenamefont {Ikeda}\ \emph {et~al.}(2024)\citenamefont {Ikeda}, \citenamefont {Kang}, \citenamefont {Kharzeev}, \citenamefont {Qian},\ and\ \citenamefont {Zhao}}]{Ikeda:2024rzv}%
  \BibitemOpen
  \bibfield  {author} {\bibinfo {author} {\bibfnamefont {K.}~\bibnamefont {Ikeda}}, \bibinfo {author} {\bibfnamefont {Z.-B.}\ \bibnamefont {Kang}}, \bibinfo {author} {\bibfnamefont {D.~E.}\ \bibnamefont {Kharzeev}}, \bibinfo {author} {\bibfnamefont {W.}~\bibnamefont {Qian}},\ and\ \bibinfo {author} {\bibfnamefont {F.}~\bibnamefont {Zhao}},\ }\bibfield  {title} {\bibinfo {title} {{Real-time chiral dynamics at finite temperature from quantum simulation}},\ }\href {https://doi.org/10.1007/JHEP10(2024)031} {\bibfield  {journal} {\bibinfo  {journal} {JHEP}\ }\textbf {\bibinfo {volume} {10}},\ \bibinfo {pages} {031}},\ \Eprint {https://arxiv.org/abs/2407.21496} {arXiv:2407.21496 [hep-ph]} \BibitemShut {NoStop}%
\bibitem [{\citenamefont {Motta}\ \emph {et~al.}(2019)\citenamefont {Motta}, \citenamefont {Sun}, \citenamefont {Tan}, \citenamefont {Rourke}, \citenamefont {Ye}, \citenamefont {Minnich}, \citenamefont {Brand\~ao},\ and\ \citenamefont {Chan}}]{Motta:2019yya}%
  \BibitemOpen
  \bibfield  {author} {\bibinfo {author} {\bibfnamefont {M.}~\bibnamefont {Motta}}, \bibinfo {author} {\bibfnamefont {C.}~\bibnamefont {Sun}}, \bibinfo {author} {\bibfnamefont {A.~T.~K.}\ \bibnamefont {Tan}}, \bibinfo {author} {\bibfnamefont {M.~J.~O.}\ \bibnamefont {Rourke}}, \bibinfo {author} {\bibfnamefont {E.}~\bibnamefont {Ye}}, \bibinfo {author} {\bibfnamefont {A.~J.}\ \bibnamefont {Minnich}}, \bibinfo {author} {\bibfnamefont {F.~G. S.~L.}\ \bibnamefont {Brand\~ao}},\ and\ \bibinfo {author} {\bibfnamefont {G.~K.-L.}\ \bibnamefont {Chan}},\ }\bibfield  {title} {\bibinfo {title} {{Determining eigenstates and thermal states on a quantum computer using quantum imaginary time evolution}},\ }\href {https://doi.org/10.1038/s41567-019-0704-4} {\bibfield  {journal} {\bibinfo  {journal} {Nature Phys.}\ }\textbf {\bibinfo {volume} {16}},\ \bibinfo {pages} {205} (\bibinfo {year} {2019})},\ \Eprint {https://arxiv.org/abs/1901.07653} {arXiv:1901.07653 [quant-ph]} \BibitemShut {NoStop}%
\bibitem [{\citenamefont {Kogut}(1983)}]{RevModPhys.55.775}%
  \BibitemOpen
  \bibfield  {author} {\bibinfo {author} {\bibfnamefont {J.~B.}\ \bibnamefont {Kogut}},\ }\bibfield  {title} {\bibinfo {title} {The lattice gauge theory approach to quantum chromodynamics},\ }\href {https://doi.org/10.1103/RevModPhys.55.775} {\bibfield  {journal} {\bibinfo  {journal} {Rev. Mod. Phys.}\ }\textbf {\bibinfo {volume} {55}},\ \bibinfo {pages} {775} (\bibinfo {year} {1983})}\BibitemShut {NoStop}%
\bibitem [{\citenamefont {'t~Hooft}(1998)}]{tHooft:1998ifg}%
  \BibitemOpen
  \bibfield  {author} {\bibinfo {author} {\bibfnamefont {G.}~\bibnamefont {'t~Hooft}},\ }\bibfield  {title} {\bibinfo {title} {{Topological aspects of quantum chromodynamics}},\ }in\ \href@noop {} {\emph {\bibinfo {booktitle} {{International School of Nuclear Physics: 20th Course: Heavy Ion Collisions from Nuclear to Quark Matter (Erice 98)}}}}\ (\bibinfo {year} {1998})\ pp.\ \bibinfo {pages} {216--236},\ \Eprint {https://arxiv.org/abs/hep-th/9812204} {arXiv:hep-th/9812204} \BibitemShut {NoStop}%
\bibitem [{\citenamefont {Atas}\ \emph {et~al.}(2021)\citenamefont {Atas}, \citenamefont {Zhang}, \citenamefont {Lewis}, \citenamefont {Jahanpour}, \citenamefont {Haase},\ and\ \citenamefont {Muschik}}]{Atas:2021ext}%
  \BibitemOpen
  \bibfield  {author} {\bibinfo {author} {\bibfnamefont {Y.~Y.}\ \bibnamefont {Atas}}, \bibinfo {author} {\bibfnamefont {J.}~\bibnamefont {Zhang}}, \bibinfo {author} {\bibfnamefont {R.}~\bibnamefont {Lewis}}, \bibinfo {author} {\bibfnamefont {A.}~\bibnamefont {Jahanpour}}, \bibinfo {author} {\bibfnamefont {J.~F.}\ \bibnamefont {Haase}},\ and\ \bibinfo {author} {\bibfnamefont {C.~A.}\ \bibnamefont {Muschik}},\ }\bibfield  {title} {\bibinfo {title} {{SU(2) hadrons on a quantum computer via a variational approach}},\ }\href {https://doi.org/10.1038/s41467-021-26825-4} {\bibfield  {journal} {\bibinfo  {journal} {Nature Commun.}\ }\textbf {\bibinfo {volume} {12}},\ \bibinfo {pages} {6499} (\bibinfo {year} {2021})},\ \Eprint {https://arxiv.org/abs/2102.08920} {arXiv:2102.08920 [quant-ph]} \BibitemShut {NoStop}%
\bibitem [{\citenamefont {Ale}\ \emph {et~al.}(2024)\citenamefont {Ale}, \citenamefont {Bauer}, \citenamefont {Jha}, \citenamefont {Ringer},\ and\ \citenamefont {Siopsis}}]{Ale:2024uxf}%
  \BibitemOpen
  \bibfield  {author} {\bibinfo {author} {\bibfnamefont {V.}~\bibnamefont {Ale}}, \bibinfo {author} {\bibfnamefont {N.~M.}\ \bibnamefont {Bauer}}, \bibinfo {author} {\bibfnamefont {R.~G.}\ \bibnamefont {Jha}}, \bibinfo {author} {\bibfnamefont {F.}~\bibnamefont {Ringer}},\ and\ \bibinfo {author} {\bibfnamefont {G.}~\bibnamefont {Siopsis}},\ }\bibfield  {title} {\bibinfo {title} {{Quantum computation of SU(2) lattice gauge theory with continuous variables}},\ }\href@noop {} {\  (\bibinfo {year} {2024})},\ \Eprint {https://arxiv.org/abs/2410.14580} {arXiv:2410.14580 [hep-lat]} \BibitemShut {NoStop}%
\bibitem [{\citenamefont {Turro}\ \emph {et~al.}(2024)\citenamefont {Turro}, \citenamefont {Ciavarella},\ and\ \citenamefont {Yao}}]{Turro:2024pxu}%
  \BibitemOpen
  \bibfield  {author} {\bibinfo {author} {\bibfnamefont {F.}~\bibnamefont {Turro}}, \bibinfo {author} {\bibfnamefont {A.}~\bibnamefont {Ciavarella}},\ and\ \bibinfo {author} {\bibfnamefont {X.}~\bibnamefont {Yao}},\ }\bibfield  {title} {\bibinfo {title} {{Classical and quantum computing of shear viscosity for (2+1)D SU(2) gauge theory}},\ }\href {https://doi.org/10.1103/PhysRevD.109.114511} {\bibfield  {journal} {\bibinfo  {journal} {Phys. Rev. D}\ }\textbf {\bibinfo {volume} {109}},\ \bibinfo {pages} {114511} (\bibinfo {year} {2024})},\ \Eprint {https://arxiv.org/abs/2402.04221} {arXiv:2402.04221 [hep-lat]} \BibitemShut {NoStop}%
\bibitem [{\citenamefont {Farhi}\ \emph {et~al.}(2014)\citenamefont {Farhi}, \citenamefont {Goldstone},\ and\ \citenamefont {Gutmann}}]{Farhi:2014ych}%
  \BibitemOpen
  \bibfield  {author} {\bibinfo {author} {\bibfnamefont {E.}~\bibnamefont {Farhi}}, \bibinfo {author} {\bibfnamefont {J.}~\bibnamefont {Goldstone}},\ and\ \bibinfo {author} {\bibfnamefont {S.}~\bibnamefont {Gutmann}},\ }\bibfield  {title} {\bibinfo {title} {{A Quantum Approximate Optimization Algorithm}},\ }\href@noop {} {\  (\bibinfo {year} {2014})},\ \Eprint {https://arxiv.org/abs/1411.4028} {arXiv:1411.4028 [quant-ph]} \BibitemShut {NoStop}%
\bibitem [{\citenamefont {Taylor}(1979)}]{Taylor1979GaugeTO}%
  \BibitemOpen
  \bibfield  {author} {\bibinfo {author} {\bibfnamefont {J.~C.}\ \bibnamefont {Taylor}},\ }\href@noop {} {\emph {\bibinfo {title} {{Gauge Theories of Weak Interactions}}}},\ Cambridge Monographs on Mathematical Physics\ (\bibinfo  {publisher} {Cambridge Univ. Press},\ \bibinfo {address} {Cambridge, UK},\ \bibinfo {year} {1979})\BibitemShut {NoStop}%
\bibitem [{\citenamefont {Kogut}\ and\ \citenamefont {Susskind}(1975)}]{Kogut:1974ag}%
  \BibitemOpen
  \bibfield  {author} {\bibinfo {author} {\bibfnamefont {J.~B.}\ \bibnamefont {Kogut}}\ and\ \bibinfo {author} {\bibfnamefont {L.}~\bibnamefont {Susskind}},\ }\bibfield  {title} {\bibinfo {title} {{Hamiltonian Formulation of Wilson's Lattice Gauge Theories}},\ }\href {https://doi.org/10.1103/PhysRevD.11.395} {\bibfield  {journal} {\bibinfo  {journal} {Phys. Rev. D}\ }\textbf {\bibinfo {volume} {11}},\ \bibinfo {pages} {395} (\bibinfo {year} {1975})}\BibitemShut {NoStop}%
\bibitem [{\citenamefont {Zohar}\ and\ \citenamefont {Burrello}(2015)}]{Zohar:2014qma}%
  \BibitemOpen
  \bibfield  {author} {\bibinfo {author} {\bibfnamefont {E.}~\bibnamefont {Zohar}}\ and\ \bibinfo {author} {\bibfnamefont {M.}~\bibnamefont {Burrello}},\ }\bibfield  {title} {\bibinfo {title} {{Formulation of lattice gauge theories for quantum simulations}},\ }\href {https://doi.org/10.1103/PhysRevD.91.054506} {\bibfield  {journal} {\bibinfo  {journal} {Phys. Rev. D}\ }\textbf {\bibinfo {volume} {91}},\ \bibinfo {pages} {054506} (\bibinfo {year} {2015})},\ \Eprint {https://arxiv.org/abs/1409.3085} {arXiv:1409.3085 [quant-ph]} \BibitemShut {NoStop}%
\bibitem [{\citenamefont {Grabowska}\ \emph {et~al.}(2024)\citenamefont {Grabowska}, \citenamefont {Kane},\ and\ \citenamefont {Bauer}}]{Grabowska:2024emw}%
  \BibitemOpen
  \bibfield  {author} {\bibinfo {author} {\bibfnamefont {D.~M.}\ \bibnamefont {Grabowska}}, \bibinfo {author} {\bibfnamefont {C.~F.}\ \bibnamefont {Kane}},\ and\ \bibinfo {author} {\bibfnamefont {C.~W.}\ \bibnamefont {Bauer}},\ }\bibfield  {title} {\bibinfo {title} {{A Fully Gauge-Fixed SU(2) Hamiltonian for Quantum Simulations}},\ }\href@noop {} {\  (\bibinfo {year} {2024})},\ \Eprint {https://arxiv.org/abs/2409.10610} {arXiv:2409.10610 [quant-ph]} \BibitemShut {NoStop}%
\bibitem [{\citenamefont {Hamer}(1977)}]{Hamer:1976bj}%
  \BibitemOpen
  \bibfield  {author} {\bibinfo {author} {\bibfnamefont {C.~J.}\ \bibnamefont {Hamer}},\ }\bibfield  {title} {\bibinfo {title} {{Lattice Model Calculations for SU(2) Yang-Mills Theory in (1+1)-Dimensions}},\ }\href {https://doi.org/10.1016/0550-3213(77)90334-0} {\bibfield  {journal} {\bibinfo  {journal} {Nucl. Phys. B}\ }\textbf {\bibinfo {volume} {121}},\ \bibinfo {pages} {159} (\bibinfo {year} {1977})},\ \bibinfo {note} {[Addendum: Nucl.Phys.B 132, 542 (1978)]}\BibitemShut {NoStop}%
\bibitem [{\citenamefont {Bringoltz}(2009)}]{Bringoltz:2008iu}%
  \BibitemOpen
  \bibfield  {author} {\bibinfo {author} {\bibfnamefont {B.}~\bibnamefont {Bringoltz}},\ }\bibfield  {title} {\bibinfo {title} {{Volume dependence of two-dimensional large-N QCD with a nonzero density of baryons}},\ }\href {https://doi.org/10.1103/PhysRevD.79.105021} {\bibfield  {journal} {\bibinfo  {journal} {Phys. Rev. D}\ }\textbf {\bibinfo {volume} {79}},\ \bibinfo {pages} {105021} (\bibinfo {year} {2009})},\ \Eprint {https://arxiv.org/abs/0811.4141} {arXiv:0811.4141 [hep-lat]} \BibitemShut {NoStop}%
\bibitem [{\citenamefont {Lenz}\ \emph {et~al.}(1994)\citenamefont {Lenz}, \citenamefont {Naus},\ and\ \citenamefont {Thies}}]{Lenz:1994cv}%
  \BibitemOpen
  \bibfield  {author} {\bibinfo {author} {\bibfnamefont {F.}~\bibnamefont {Lenz}}, \bibinfo {author} {\bibfnamefont {H.~W.~L.}\ \bibnamefont {Naus}},\ and\ \bibinfo {author} {\bibfnamefont {M.}~\bibnamefont {Thies}},\ }\bibfield  {title} {\bibinfo {title} {{QCD in the axial gauge representation}},\ }\href {https://doi.org/10.1006/aphy.1994.1071} {\bibfield  {journal} {\bibinfo  {journal} {Annals Phys.}\ }\textbf {\bibinfo {volume} {233}},\ \bibinfo {pages} {317} (\bibinfo {year} {1994})}\BibitemShut {NoStop}%
\bibitem [{\citenamefont {Sala}\ \emph {et~al.}(2018)\citenamefont {Sala}, \citenamefont {Shi}, \citenamefont {K\"uhn}, \citenamefont {Ba\~nuls}, \citenamefont {Demler},\ and\ \citenamefont {Cirac}}]{Sala:2018dui}%
  \BibitemOpen
  \bibfield  {author} {\bibinfo {author} {\bibfnamefont {P.}~\bibnamefont {Sala}}, \bibinfo {author} {\bibfnamefont {T.}~\bibnamefont {Shi}}, \bibinfo {author} {\bibfnamefont {S.}~\bibnamefont {K\"uhn}}, \bibinfo {author} {\bibfnamefont {M.~C.}\ \bibnamefont {Ba\~nuls}}, \bibinfo {author} {\bibfnamefont {E.}~\bibnamefont {Demler}},\ and\ \bibinfo {author} {\bibfnamefont {J.~I.}\ \bibnamefont {Cirac}},\ }\bibfield  {title} {\bibinfo {title} {{Variational study of U(1) and SU(2) lattice gauge theories with Gaussian states in 1+1 dimensions}},\ }\href {https://doi.org/10.1103/PhysRevD.98.034505} {\bibfield  {journal} {\bibinfo  {journal} {Phys. Rev. D}\ }\textbf {\bibinfo {volume} {98}},\ \bibinfo {pages} {034505} (\bibinfo {year} {2018})},\ \Eprint {https://arxiv.org/abs/1805.05190} {arXiv:1805.05190 [hep-lat]} \BibitemShut {NoStop}%
\bibitem [{\citenamefont {Jordan}\ and\ \citenamefont {Wigner}(1928)}]{Jordan1928berDP}%
  \BibitemOpen
  \bibfield  {author} {\bibinfo {author} {\bibfnamefont {P.}~\bibnamefont {Jordan}}\ and\ \bibinfo {author} {\bibfnamefont {E.}~\bibnamefont {Wigner}},\ }\bibfield  {title} {\bibinfo {title} {{\"U}ber das paulische {\"a}quivalenzverbot},\ }\href {https://api.semanticscholar.org/CorpusID:126400679} {\bibfield  {journal} {\bibinfo  {journal} {Zeitschrift f{\"u}r Physik}\ }\textbf {\bibinfo {volume} {47}},\ \bibinfo {pages} {631} (\bibinfo {year} {1928})}\BibitemShut {NoStop}%
\bibitem [{\citenamefont {Lowdin}(1955{\natexlab{a}})}]{Lowdin:1955zzb}%
  \BibitemOpen
  \bibfield  {author} {\bibinfo {author} {\bibfnamefont {P.-O.}\ \bibnamefont {Lowdin}},\ }\bibfield  {title} {\bibinfo {title} {{Quantum Theory of Many-Particle Systems. 1. Physical Interpretations by Means of Density Matrices, Natural Spin-Orbitals, and Convergence Problems in the Method of Configurational Interaction}},\ }\href {https://doi.org/10.1103/PhysRev.97.1474} {\bibfield  {journal} {\bibinfo  {journal} {Phys. Rev.}\ }\textbf {\bibinfo {volume} {97}},\ \bibinfo {pages} {1474} (\bibinfo {year} {1955}{\natexlab{a}})}\BibitemShut {NoStop}%
\bibitem [{\citenamefont {Lowdin}(1955{\natexlab{b}})}]{Lowdin:1955zza}%
  \BibitemOpen
  \bibfield  {author} {\bibinfo {author} {\bibfnamefont {P.-O.}\ \bibnamefont {Lowdin}},\ }\bibfield  {title} {\bibinfo {title} {{Quantum Theory of Many-Particle Systems. 2. Study of the Ordinary Hartree-Fock Approximation}},\ }\href {https://doi.org/10.1103/PhysRev.97.1490} {\bibfield  {journal} {\bibinfo  {journal} {Phys. Rev.}\ }\textbf {\bibinfo {volume} {97}},\ \bibinfo {pages} {1490} (\bibinfo {year} {1955}{\natexlab{b}})}\BibitemShut {NoStop}%
\bibitem [{\citenamefont {Lowdin}(1955{\natexlab{c}})}]{Lowdin:1955zz}%
  \BibitemOpen
  \bibfield  {author} {\bibinfo {author} {\bibfnamefont {P.-O.}\ \bibnamefont {Lowdin}},\ }\bibfield  {title} {\bibinfo {title} {{Quantum Theory of Many-Particle Systems. 3. Extension of the Hartree-Fock Scheme to Include Degenerate Systems and Correlation Effects}},\ }\href {https://doi.org/10.1103/PhysRev.97.1509} {\bibfield  {journal} {\bibinfo  {journal} {Phys. Rev.}\ }\textbf {\bibinfo {volume} {97}},\ \bibinfo {pages} {1509} (\bibinfo {year} {1955}{\natexlab{c}})}\BibitemShut {NoStop}%
\bibitem [{\citenamefont {Bogolubov}\ and\ \citenamefont {Bogolubov}(2009)}]{10.5555/1823040}%
  \BibitemOpen
  \bibfield  {author} {\bibinfo {author} {\bibfnamefont {N.~N.}\ \bibnamefont {Bogolubov}}\ and\ \bibinfo {author} {\bibfnamefont {N.~N.}\ \bibnamefont {Bogolubov}, \bibfnamefont {Jr.}},\ }\href@noop {} {\emph {\bibinfo {title} {Introduction to Quantum Statistical Mechanics}}},\ \bibinfo {edition} {2nd}\ ed.\ (\bibinfo  {publisher} {World Scientific Publishing Co., Inc.},\ \bibinfo {address} {USA},\ \bibinfo {year} {2009})\BibitemShut {NoStop}%
\bibitem [{\citenamefont {Verdon}\ \emph {et~al.}(2019)\citenamefont {Verdon}, \citenamefont {Marks}, \citenamefont {Nanda}, \citenamefont {Leichenauer},\ and\ \citenamefont {Hidary}}]{Verdon:2019hpy}%
  \BibitemOpen
  \bibfield  {author} {\bibinfo {author} {\bibfnamefont {G.}~\bibnamefont {Verdon}}, \bibinfo {author} {\bibfnamefont {J.}~\bibnamefont {Marks}}, \bibinfo {author} {\bibfnamefont {S.}~\bibnamefont {Nanda}}, \bibinfo {author} {\bibfnamefont {S.}~\bibnamefont {Leichenauer}},\ and\ \bibinfo {author} {\bibfnamefont {J.}~\bibnamefont {Hidary}},\ }\bibfield  {title} {\bibinfo {title} {{Quantum Hamiltonian-Based Models and the Variational Quantum Thermalizer Algorithm}},\ }\href@noop {} {\  (\bibinfo {year} {2019})},\ \Eprint {https://arxiv.org/abs/1910.02071} {arXiv:1910.02071 [quant-ph]} \BibitemShut {NoStop}%
\bibitem [{\citenamefont {Wiersema}\ \emph {et~al.}(2020)\citenamefont {Wiersema}, \citenamefont {Zhou}, \citenamefont {de~Sereville}, \citenamefont {Carrasquilla}, \citenamefont {Kim},\ and\ \citenamefont {Yuen}}]{Wiersema:2020ipa}%
  \BibitemOpen
  \bibfield  {author} {\bibinfo {author} {\bibfnamefont {R.}~\bibnamefont {Wiersema}}, \bibinfo {author} {\bibfnamefont {C.}~\bibnamefont {Zhou}}, \bibinfo {author} {\bibfnamefont {Y.}~\bibnamefont {de~Sereville}}, \bibinfo {author} {\bibfnamefont {J.~F.}\ \bibnamefont {Carrasquilla}}, \bibinfo {author} {\bibfnamefont {Y.~B.}\ \bibnamefont {Kim}},\ and\ \bibinfo {author} {\bibfnamefont {H.}~\bibnamefont {Yuen}},\ }\bibfield  {title} {\bibinfo {title} {{Exploring Entanglement and Optimization within the Hamiltonian Variational Ansatz}},\ }\href {https://doi.org/10.1103/PRXQuantum.1.020319} {\bibfield  {journal} {\bibinfo  {journal} {PRX Quantum}\ }\textbf {\bibinfo {volume} {1}},\ \bibinfo {pages} {020319} (\bibinfo {year} {2020})}\BibitemShut {NoStop}%
\bibitem [{\citenamefont {LeCun}\ \emph {et~al.}(2012)\citenamefont {LeCun}, \citenamefont {Bottou}, \citenamefont {Orr},\ and\ \citenamefont {Muller}}]{8c8eccbbe8a040118afa8f8423da1fe2}%
  \BibitemOpen
  \bibfield  {author} {\bibinfo {author} {\bibfnamefont {Y.}~\bibnamefont {LeCun}}, \bibinfo {author} {\bibfnamefont {L.}~\bibnamefont {Bottou}}, \bibinfo {author} {\bibfnamefont {G.}~\bibnamefont {Orr}},\ and\ \bibinfo {author} {\bibfnamefont {K.}~\bibnamefont {Muller}},\ }\bibinfo {title} {Efficient backprop},\ in\ \href {https://doi.org/10.1007/978-3-642-35289-8_3} {\emph {\bibinfo {booktitle} {Neural Networks}}},\ \bibinfo {series and number} {Lecture Notes in Computer Science (including subseries Lecture Notes in Artificial Intelligence and Lecture Notes in Bioinformatics)}\ (\bibinfo  {publisher} {Springer Verlag},\ \bibinfo {year} {2012})\ pp.\ \bibinfo {pages} {9--48},\ \bibinfo {note} {copyright: Copyright 2021 Elsevier B.V., All rights reserved.}\BibitemShut {Stop}%
\bibitem [{\citenamefont {Weinberg}\ and\ \citenamefont {Bukov}(2017)}]{Weinberg:2017igw}%
  \BibitemOpen
  \bibfield  {author} {\bibinfo {author} {\bibfnamefont {P.}~\bibnamefont {Weinberg}}\ and\ \bibinfo {author} {\bibfnamefont {M.}~\bibnamefont {Bukov}},\ }\bibfield  {title} {\bibinfo {title} {{QuSpin: a Python package for dynamics and exact diagonalisation of quantum many body systems part I: spin chains}},\ }\href {https://doi.org/10.21468/SciPostPhys.2.1.003} {\bibfield  {journal} {\bibinfo  {journal} {SciPost Phys.}\ }\textbf {\bibinfo {volume} {2}},\ \bibinfo {pages} {003} (\bibinfo {year} {2017})},\ \Eprint {https://arxiv.org/abs/1610.03042} {arXiv:1610.03042} \BibitemShut {NoStop}%
\bibitem [{\citenamefont {Weinberg}\ and\ \citenamefont {Bukov}(2019)}]{Weinberg:2019rfm}%
  \BibitemOpen
  \bibfield  {author} {\bibinfo {author} {\bibfnamefont {P.}~\bibnamefont {Weinberg}}\ and\ \bibinfo {author} {\bibfnamefont {M.}~\bibnamefont {Bukov}},\ }\bibfield  {title} {\bibinfo {title} {{QuSpin: a Python package for dynamics and exact diagonalisation of quantum many body systems. Part II: bosons, fermions and higher spins}},\ }\href {https://doi.org/10.21468/SciPostPhys.7.2.020} {\bibfield  {journal} {\bibinfo  {journal} {SciPost Phys.}\ }\textbf {\bibinfo {volume} {7}},\ \bibinfo {pages} {020} (\bibinfo {year} {2019})},\ \Eprint {https://arxiv.org/abs/1804.06782} {arXiv:1804.06782} \BibitemShut {NoStop}%
\bibitem [{\citenamefont {Javadi-Abhari}\ \emph {et~al.}(2024)\citenamefont {Javadi-Abhari} \emph {et~al.}}]{Javadi-Abhari:2024kbf}%
  \BibitemOpen
  \bibfield  {author} {\bibinfo {author} {\bibfnamefont {A.}~\bibnamefont {Javadi-Abhari}} \emph {et~al.},\ }\bibfield  {title} {\bibinfo {title} {{Quantum computing with Qiskit}},\ }\href@noop {} {\  (\bibinfo {year} {2024})},\ \Eprint {https://arxiv.org/abs/2405.08810} {arXiv:2405.08810 [quant-ph]} \BibitemShut {NoStop}%
\bibitem [{\citenamefont {Chelabi}\ \emph {et~al.}(2016)\citenamefont {Chelabi}, \citenamefont {Fang}, \citenamefont {Huang}, \citenamefont {Li},\ and\ \citenamefont {Wu}}]{Chelabi:2015cwn}%
  \BibitemOpen
  \bibfield  {author} {\bibinfo {author} {\bibfnamefont {K.}~\bibnamefont {Chelabi}}, \bibinfo {author} {\bibfnamefont {Z.}~\bibnamefont {Fang}}, \bibinfo {author} {\bibfnamefont {M.}~\bibnamefont {Huang}}, \bibinfo {author} {\bibfnamefont {D.}~\bibnamefont {Li}},\ and\ \bibinfo {author} {\bibfnamefont {Y.-L.}\ \bibnamefont {Wu}},\ }\bibfield  {title} {\bibinfo {title} {{Realization of chiral symmetry breaking and restoration in holographic QCD}},\ }\href {https://doi.org/10.1103/PhysRevD.93.101901} {\bibfield  {journal} {\bibinfo  {journal} {Phys. Rev. D}\ }\textbf {\bibinfo {volume} {93}},\ \bibinfo {pages} {101901} (\bibinfo {year} {2016})},\ \Eprint {https://arxiv.org/abs/1511.02721} {arXiv:1511.02721 [hep-ph]} \BibitemShut {NoStop}%
\bibitem [{\citenamefont {Wetterich}(2002)}]{Wetterich:2001dq}%
  \BibitemOpen
  \bibfield  {author} {\bibinfo {author} {\bibfnamefont {C.}~\bibnamefont {Wetterich}},\ }\bibfield  {title} {\bibinfo {title} {{Connection between chiral symmetry restoration and deconfinement}},\ }\href {https://doi.org/10.1103/PhysRevD.66.056003} {\bibfield  {journal} {\bibinfo  {journal} {Phys. Rev. D}\ }\textbf {\bibinfo {volume} {66}},\ \bibinfo {pages} {056003} (\bibinfo {year} {2002})},\ \Eprint {https://arxiv.org/abs/hep-ph/0102044} {arXiv:hep-ph/0102044} \BibitemShut {NoStop}%
\end{thebibliography}%

\end{document}